\documentclass[a4paper,UKenglish,cleveref, autoref,dvipsnames]{lipics-v2021}
\pdfoutput=1

\usepackage{xspace,xargs,xcolor}

\usepackage{amsmath,amssymb,tabularx,hyperref}
\usepackage{colortbl}

\usepackage{tikz} 
\usetikzlibrary{tikzmark}

\hideLIPIcs

\newcommand{\vir}[1]{``#1''}

\newcommand{\Run}{\rho}
\newcommand{\runmin}{Opt}
\newcommand{\bw}{bw}

\newcommand{\BWT}{BWT\xspace}
\newcommand{\BWTs}{BWTs\xspace}
\newcommand{\rindex}{$r$-index\xspace}

\newcommand{\bwt}{\ensuremath{bwt}}
\newcommand{\ABWT}{ABWT\xspace}

\newcommand{\rank}{\ensuremath{\mathsf{rank}}\xspace}
\newcommand{\sel}{\ensuremath{\mathsf{select}}\xspace}
\newcommand{\access}{\ensuremath{\mathsf{access}}\xspace}

\newcommand{\pos}{\ensuremath{\mathsf{pos}}}

\newcommand{\Rng}[1]{R[#1]}
\newcommand{\Rngx}[1]{R^{\ast}[#1]}
\newcommand{\Mx}{M_{*}}
\newcommand{\BWTx}{BWT_{*}}
\newcommand{\asize}{{\sigma}}
\newcommand{\Alpha}{\Sigma}
\newcommand{\Oh}{{\cal O}}

\newcommand{\xchar}{{c}}
\newcommand{\bone}{{\bf 1}}
\newcommand{\btwo}{{\bf 2}}
\newcommand{\bthree}{{\bf 3}}

\long\def\ignore#1{}

\definecolor{darkpastelgreen}{rgb}{0.01, 0.5, 0.24}
\newcommand{\darkgreen}{\textcolor{darkpastelgreen}}

\definecolor{cerisepink}{rgb}{0.93, 0.23, 0.51}

\definecolor{newredv}{rgb}{0.6,0.01,0.24}

\long\def\ignore#1{\vskip 0pt}

\nolinenumbers

\title{A New Class of String Transformations for Compressed Text Indexing}

\author{Raffaele Giancarlo}{University of Palermo, Dipartimento di Matematica e Informatica, Italy}{raffaele.giancarlo@unipa.it}{0000-0002-6286-8871}{}

\author{Giovanni Manzini}{University of Pisa, Dipartimento di Informatica, Italy}{giovanni.manzini@unipi.it}{0000-0002-5047-0196}{}

\author{Antonio Restivo}{University of Palermo, Dipartimento di Matematica e Informatica, Italy}{antonio.restivo@unipa.it}{}{}

\author{Giovanna Rosone}{University of Pisa, Dipartimento di Informatica, Italy}{giovanna.rosone@unipi.it}{0000-0001-5075-1214}{}

\author{Marinella Sciortino}{University of Palermo, Dipartimento di Matematica e Informatica, Italy}{marinella.sciortino@unipa.it}{0000-0001-6928-0168}{}

\authorrunning{R. Giancarlo and G. Manzini and A. Restivo and G. Rosone and M. Sciortino}

\Copyright{Raffaele Giancarlo and Giovanni Manzini and Antonio Restivo and Giovanna Rosone and Marinella Sciortino}

\long\def\ignore#1{\vskip 0pt}


\funding{
GM and GR are partially supported by PNRR ECS00000017 Tuscany Health Ecosystem, Spoke 6 ``Precision medicine \& personalized healthcare'', funded by the European Commission under the NextGeneration EU programme.
GM is partially supported by the ICSC – Centro Nazionale di Ricerca in High-Performance Computing, Big Data and Quantum Computing, Spoke 1
``FutureHPC \& BigData'' funded by European Union, NextGeneration EU programme.
MS is partially supported by the PNRR project Ita\-lian Strengthening of ESFRI RI RESILIENCE (ITSERR) funded by the European Union, NextGeneration EU programme, CUP~B53C22001770006 and by ``FFR-Sciortino''.
GR and MS are partially supported by INdAM-GNCS Project 2022 \vir{Metodi combinatori per indicizzare e confrontare collezioni di testi altamente ripetitivi}, CUP E55F22000270001. 
GM, GR and MS are partially supported by INdAM-GNCS Project 2023 \vir{Linguaggi regolari ordinati, indicizzazione e confronto di collezioni di stringhe, con applicazioni}, CUP E53C22001930001.
RG is partially supported by INDAM-GNCS Project 2023 \vir{Approcci computazionali per il supporto alle decisioni nella Medicina di Precisione}, CUP E53C22001930001.
RG and GM are partially supported by MIUR-PRIN project \vir{Multicriteria Data Structures and Algorithms: from compressed to learned indexes, and beyond} grant n.~2017WR7SHH.}

\begin{document}





\maketitle

\begin{abstract}

Introduced about thirty years ago in the field of data compression, the Burrows-Wheeler Transform (BWT) is a string transformation that, besides being a \emph{booster} of the performance of memoryless compressors, plays a fundamental role in the design of efficient self-indexing compressed data structures. Finding other string transformations {with the same remarkable properties of BWT} has been a challenge for many researchers for a long time. Among the known BWT variants, the only one that has been recently shown to be a valid alternative to BWT is the \emph{Alternating BWT} (ABWT), an invertible string transformation introduced about ten years ago in connection with a generalization of Lyndon words. In this paper, we introduce a whole class of {new} string transformations, called \emph{local orderings-based transformations}, which have all the \vir{myriad virtues} of BWT. We show that this new family is a special case of a much larger class of transformations, based on \emph{context adaptive alphabet orderings}, that includes BWT and ABWT.
{Although all transformations support pattern search,}
we show that, in the general case, the transformations within our larger class may take quadratic time for inversion and pattern search. As a further result, we show that the local orderings-based transformations can be used for the construction of the recently introduced $r$-index, which makes them suitable also for highly repetitive collections.
In this context, we consider the problem of finding, for a given string, the BWT variant that minimizes the number of runs in the transformed string, and we provide an algorithm solving this problem in linear time.

\end{abstract}

\section{Introduction}

Michael Burrows and David Wheeler introduced in 1994 a reversible word transformation~\cite{bwt94}, denoted by $\BWT$, that turned out to have \vir{myriad virtues}. 
At the time of its introduction in the field of text compression, the Burrows-Wheeler Transform was perceived as a {magic box}: when used as a preprocessing step it would bring rather weak compressors to be competitive in terms of compression ratio with the best ones available~\cite{cj/fenwick96}. In the years that followed, many studies have shown the effectiveness of \BWT and its central role in the field of data compression, due to the fact that it can be seen as a \vir{booster} of the performance of memoryless compressors~\cite{FGMS2005,GIA07,Manzini2001}.

The importance of this transformation has been further increased when in~\cite{Ferragina:2000} it was proven that, in addition to making easier to represent a string in space close to its entropy, it also makes easier to search for pattern occurrences in the original string. After this discovery, data transformations inspired by the \BWT have been proposed for compactly representing and searching other combinatorial objects such as: trees, graphs, finite automata, and even string alignments. See~\cite{iandc/AlankoDPP21,tcs/GagieMS17} for an attempt to unify some of these results and~\cite{Nav16} for an in-depth treatment of the field of compact data structures.

Going back to the original Burrows-Wheeler string transformation, we can summarize its salient features as follows: \bone) it can be  computed and inverted in linear time, \btwo) it produces strings which are provably compressible in terms of the high order entropy of the input, \bthree) it supports pattern search directly on the transformed string in time proportional to the pattern length. It is the {\em combination} of these three properties that makes the \BWT a fundamental tool for the design of compressed self-indices. In Section~\ref{sec:prel} we review these properties and also the many attempts to modify the original design. However, we recall that, despite more than twenty years of intense scrutiny, the only non trivial known \BWT variant that fully satisfies properties \bone--\bthree\ is the {\em Alternating \BWT} (\ABWT). The \ABWT has been introduced in~\cite{GesselRestivoReutenauer2012} in the field of combinatorics of words and its basic algorithmic properties have been described in~\cite{dlt/GiancarloMRRS18,abwt_jou}.

In this paper we introduce a new {\em whole family} of transformations that satisfy properties \bone--\bthree\ and can therefore replace the \BWT in the construction of compressed self-indices with the same time efficiency of the original \BWT and the potential of achieving better compression. We show that our family, supporting linear time computation, inversion, and search, is a special case of a much larger class of transformations that also satisfy properties \bone--\bthree\ except that, in the general case, inversion and pattern search may take quadratic time. Our larger class includes as special cases also the \BWT and the \ABWT and therefore it constitutes a natural candidate for the study of additional properties shared by all known \BWT variants.

\ignore{This class was mentioned in~\cite{FGMS2005}, where it was observed that these transformations can be computed in linear time and that they produce highly compressible strings (in the sense of property \btwo).  
However, in that paper, the problem of the invertibility of these transformations was left open: here we prove that they are all invertible by providing a quadratic time inversion algorithm. In addition, we show that these transformations support pattern search directly in the transformed text in time quadratic with the pattern length.}

More in detail, in Section~\ref{sec:genBWT} we describe a class of string transformations based on {\em context adaptive alphabet orderings}. 
The main feature of the above class of transformations is that, in the rotation sorting phase, we use alphabet orderings that depend on the context (i.e., the longest common prefix of the rotations being compared). 
{We prove that such transformations are always invertible and provide a general inversion algorithm that runs in time quadratic with respect to the string length. We consider also some subclasses of such transformations in which the permutations associated to each prefix are defined by ``simple'' rules and we show that they have more efficient inversion algorithms.}

In Section~\ref{sec:local} we consider the subclass of transformations based on {\em local orderings}. In this subclass, the alphabet orderings only depend on a constant portion of the context. We prove that local ordering transformations can be inverted in linear time, and that pattern search in the transformed string takes time proportional to the pattern length. Thus, these transformations have the same properties \bone--\bthree\ that were so far prerogative of the \BWT and \ABWT. {We also show that it is always possible to implement an \rindex~\cite{GNPjacm19} on the top of a BWT based on a local ordering, thus making this transformation suitable also for highly repetitive collections.}

Having now at our disposal a wide class of string transformations with the same remarkable properties of the \BWT, it is natural to use them to improve \BWT-based data structures by selecting the one more suitable for the task.  In this paper we initiate this study by considering the problem of selecting the \BWT variant that minimizes the number of runs in the transformed string. 
The motivation is that data centers often store highly repetitive collections, such as genome databases, source code repositories, and versioned text collections. For such collections theoretical and practical evidence suggests that entropy underestimates the compressibility of the collection and we can obtain much better compression ratios exploiting runs of equal symbols in the \BWT~\cite{GNPjacm19,KaplanVerbin07,jcb/MakinenNSV10,MantaciRRS17,MantaciRRSV17,csur/Navarro21a,csur/Navarro21,RestivoRosoneTCS2011, FrosiniMRRS22, GuerriniLR22, FRSU_CPM23}.

As we review in Section~\ref{sec:runs:bkg}, run-minimization is a challenging and multifaceted problem: we believe the introduction of a new class of string transformations makes it even more interesting.
In Section~\ref{sec:runs} we contribute to this problem showing that, for constant size alphabet, for the most general class of transformations considered in this paper, the \BWT variant that minimizes the number of runs can be found in linear time using a dynamic programming algorithm. 
Although such result does not lead to a practical compression algorithm, such minimal number of runs constitutes a lower bound for the number of runs achievable by the other variants described in this paper and therefore constitutes a baseline for further theoretical or experimental studies.

A preliminary version of this paper appeared in~\cite{cpm/GiancarloMRS19}. In this new version we introduce and analyze new BWT variants, in particular the ones described in Sections~\ref{sec:lcplen} and~\ref{sec:pm}. We also added Section~\ref{sec:rindex} where we show that the recently introduced $r$-index~\cite{GNPjacm19} can be adapted to work with some of the proposed BWT variants. This result is particularly important since the $r$-index is particularly suited to highly repetitive collections which are relevant for the run minimization problem discussed in Section~\ref{sec:runs}. Another new result is the characterization of local ordering transformations contained in Section~\ref{sec:alt}. The new version also includes a more in-depth discussion of the existing literature: Section~\ref{sec:known} has been largely extended and Section~\ref{sec:runs:bkg} is new. Finally, we have added some carefully designed examples in order to better illustrate some subtle points of the exposition. All references have been updated, and new relevant results have been discussed in the paper.

\section{Notation and background}\label{sec:prel}

Let $\Alpha =\{c_1, c_2, \ldots, c_\asize\}$ be a finite ordered alphabet of size $\asize$ with $c_1< c_2< \cdots < c_\asize$, where $<$ denotes the standard lexicographic order. We denote by $\Alpha^*$ the set of strings over $\Alpha$.  Given a string $x=x_1 x_2 \cdots x_n \in \Alpha^*$, we denote by $|x|$ its length $n$. We use $\epsilon$ to denote the empty string. A \emph{factor} of $x$ is written as $x[i,j] = x_i \cdots x_j$ with $1\leq i \leq j \leq n$.  A factor of type $x[1,j]$ is called a \emph{prefix}, while a factor of type $x[i,n]$ is called a \emph{suffix}. The $i$-th symbol in $x$ is denoted by $x[i]$.  Two strings $x,y\in \Sigma^*$ are called {\em conjugate}, if $x=uv$ and $y=vu$, where $u,v\in \Sigma^*$. We also say that $x$ is a {\em cyclic rotation} of $y$. A string $x$ is {\em primitive} if all its cyclic rotations are distinct.
Given a string $x$ and $c\in\Alpha$, we write $\rank_c(x,i)$ to denote the number of occurrences of $c$ in $x[1,i]$, and $\sel_c(x,j)$ to denote the position of the $j$-th $c$ in~$x$. 

Given a primitive string $s$, we consider the matrix of all its cyclic rotations sorted in lexicographic order. Note that the rotations are all distinct by the primitivity of~$s$. The last column of the matrix is called the Burrows-Wheeler Transform of the string $s$ and it is denoted by $\bwt(s)$ (see Figure~\ref{fig:ABWTx} (left)). 
It is shown in~\cite{bwt94} that $\bwt(s)$ is always a permutation of $s$, and that there exists a linear time procedure to recover $s$, given $\bwt(s)$ and the row index $I$ of $s$ in the rotations matrix (it is $I=2$ in Figure~\ref{fig:ABWTx} (left)). 
 
In this paper we follow the assumption usually done for text indexing that the last symbol of~$s$ is a unique end-of-string symbol. This guarantees the primitivity of $s$. In addition, this assumption implies that, if we compare two cyclic rotations symbol-by-symbol, they differ as soon one of them reaches the end-of-string symbol (or sooner). This property ensures that all families of transformations defined in this paper can be computed in linear time using a suffix tree as described in Section~\ref{sec:genBWT} (note that the \BWT and \ABWT can be computed in linear time also for a generic primitive string~\cite{GIA07,abwt_jou,BCLRS21_ebwt}). The reader will notice that the algorithms computing the {\it inverse} transformations provided in this paper do not make use of the unique end-of-string symbol and therefore are valid for any primitive string.

\ignore{In this paper we follow the text indexing assumption that the last symbol of $s$ is a unique end-of-string symbol. This ensures the primitivity of $s$ and also prevents matches to wrap around in $s$. With this assumption, the \BWT can be computed in $\Oh(|s|)$ time using any linear time algorithm for Suffix Array construction.}

\begin{figure}[tb]
{
{\small
$$
\begin{array}{cc@{\ \ }c@{\ \ }c@{\ \ }c@{\ \ }c@{\ \ }c@{\ \ }c@{\ \ }c@{\ \ }c@{\ \ }l@{\ \ }}
 & F & & & &          &  &      &    & L & \\
 &\downarrow& & & &&&& &\downarrow & \\
1 & a & a & a & b & a & c & a & a & b & \\ 
2 & a & a & b & a & a & a & b & a & c & \leftarrow s\\ 
3 & a & a & b & a & c & a & a & b & a & \\ 
4 & a & b & a & a & a & b & a & c & a & \\ 
5 & a & b & a & c & a & a & b & a & a & \\
6 & a & c & a & a & b & a & a & a & b & \\ 
7 & b & a & a & a & b & a & c & a & a & \\ 
8 & b & a & c & a & a & b & a & a & a &  \\ 
9 & c & a & a & b & a & a & a & b & a &  
\end{array}
\qquad
\qquad
\begin{array}{cc@{\ \ }c@{\ \ }c@{\ \ }c@{\ \ }c@{\ \ }c@{\ \ }c@{\ \ }c@{\ \ }c@{\ \ }l@{\ \ }}
 & F & & & &          &  &      &    & L  &  \\
 &\downarrow& & & &&&& &\downarrow & \\
1 & a & c & a & a & b & a & a & a & b & \\ 
2 & a & b & a & c & a & a & b & a & a & \\
3 & a & b & a & a & a & b & a & c & a & \\ 
4 & a & a & a & b & a & c & a & a & b & \\ 
5 & a & a & b & a & a & a & b & a & c & \leftarrow s\\ 
6 & a & a & b & a & c & a & a & b & a & \\ 
7 & b & a & a & a & b & a & c & a & a & \\ 
8 & b & a & c & a & a & b & a & a & a & \\ 
9 & c & a & a & b & a & a & a & b & a &  
\end{array}
$$
}
}
\caption{The original \BWT matrix for the string $s=aabaaabac$ (left), and the \ABWT matrix of cyclic rotations sorted using the alternating lexicographic order (right).
In both matrices the horizontal arrow marks the position of the original string $s$, and the last column $L$ is the output of the transformation.}\label{fig:ABWTx}
\end{figure}

\ignore{
 & c & a & a & b & a & a & a & b & a \\ 
 & a & c & a & a & b & a & a & a & b \\ 
 & b & a & c & a & a & b & a & a & a \\ 
 & a & b & a & c & a & a & b & a & a \\
 & a & a & b & a & c & a & a & b & a \\ 
 & a & a & a & b & a & c & a & a & b \\ 
 & b & a & a & a & b & a & c & a & a \\ 
 & a & b & a & a & a & b & a & c & a \\ 
 & a & a & b & a & a & a & b & a & c \\ 
 }

The \BWT has been introduced as a data compression tool: it was empirically observed that $\bwt(s)$ usually contains long runs of equal symbols. This notion was later mathematically formalized in terms of the empirical entropy of the input string~\cite{FGMS2005,Manzini2001}. For $k \geq 0$, the $k$-th order empirical entropy of a string $x$, denoted as $H_k(x)$, is a lower bound to the compression ratio of any algorithm that encodes each symbol of $x$ using a codeword that only depends on the $k$ symbols preceding it in $x$. The simplest compressors, such as Huffman coding, in which the code of a symbol does not depend on the previous symbols, typically achieve a (modest) compression bounded in terms of the zeroth-order entropy $H_0$. This class of compressors are referred to as {\em memoryless} compressors. 
More sophisticated compressors, such as Lempel-Ziv compressors and derivatives, use knowledge from the already seen part of the input to compress the incoming symbols. They are slower than memoryless compressors but they achieve a much better compression ratio, which can be usually bounded in terms of the $k$-th order entropy of the input string for a large $k$~\cite{koma00}.

It is proven in~\cite[Theorem 5.4]{FGMS2005} that the informal statement ``the output of the \BWT is {highly compressible}'' can be formally restated saying that $\bwt(s)$ can be compressed up to $H_k(s)$, for any $k>0$, using any tool able to compress up to the zeroth-order entropy. In other words, after applying the \BWT, we can achieve high order compression using a simple (and fast) memoryless compressor. This property is often referred to as the ``boosting'' property of the \BWT.
Another remarkable property of the \BWT is that it can be used to build compressed indices. It is shown in~\cite{Ferragina:2005} how to compute the number of occurrences of a pattern $x$ in $s$ in $\Oh(t_R|x|)$ time, where $t_R$ is the cost of executing a $\rank$ query over $\bwt(s)$. This result has spurred a great interest in data structures representing compactly a string $x$ and efficiently supporting the queries \rank, \sel, and \access (return $x[i]$ given $i$, which is a nontrivial operation when $x$ is represented in compressed form) and there are now many alternative solutions, with different trade-offs. In this paper, we assume a RAM model with word size $w$ and an alphabet of size $\sigma = w^{\Oh(1)}$. Under this assumption, we make use of the following result (Theorem~5.1 in~\cite{BNtalg14}):

\begin{theorem}\label{theo:rsa}
Let $s$ denote a string over an alphabet of size $\sigma= w^{\Oh(1)}$. We can represent~$s$ in $|s| H_0(s) + o(|s|)$ bits and support constant time \rank, \sel, and \access queries.
\end{theorem}

{The assumption $\sigma = w^{\Oh(1)}$ is only required to ensure that \rank, \sel, and \access take constant time; we use it to simplify the statements of our results. It is straightforward to verify that for larger alphabets our results still hold with the bounds on the running times multiplied by the largest of the cost of the above operations over the larger alphabets.}

The properties of the \BWT of being {\em compressible} and {\em searchable} combine nicely to give us {\it indexing capabilities} in {\it compressed space}. Indeed, combining a zeroth-order representation supporting \rank, \sel, and \access queries with the boosting property of the \BWT, we obtain a full text self-index for $s$ that uses space bounded by $|s|H_k(s) + o(|s|)$ bits for $k=o(\log_\sigma(n))$, see~\cite{Ferragina:2005,Mak15,Nav16,NM-survey07} for further details on these results and on the field of compressed data structures and algorithms that originated from this area of research.

\ignore{\color{red}
\begin{theorem}
Given a memoryless compressor $A$, compressing each string $x$ in up to $\lambda|x|H_0(x) + \mu |x|$, we can find in $\Oh(s)$ time a partition $x_1 x_2 \cdots x_h$ of $\bwt(s)$  
\end{theorem}}

\subsection{Known BWT variants}\label{sec:known}

We observed that the salient features of the Burrows-Wheeler transformation can be summarized as follows: \bone) it can be  computed and inverted in linear time, \btwo) it produces strings which are provably compressible in terms of the high order entropy of the input, \bthree) it supports linear time pattern search directly on the transformed string. The {\em combination} of these three properties makes the \BWT a fundamental tool for the design of compressed self-indices.  Over the years, many variants of the original \BWT have been proposed; in the following we review them, in roughly chronological order, emphasizing to what extent they share the features \bone--\bthree\ mentioned above.  

The original \BWT is defined by sorting in lexicographic order all the cyclic rotations of the input string. In~\cite{Schindler1997} Schindler proposes a {\em bounded context} transformation that differs from the \BWT in the fact that the rotations are lexicographically sorted considering only the first $\ell$ symbols of each rotation. 
In \cite{spe/CulpepperPP12,ccp/PetriNCP11} it has been shown that this variant satisfies properties \bone--\bthree, with the limitation that the compression ratio can reach at maximum the $\ell$-th order entropy and that it supports searches of patterns of length at most $\ell$. Chapin and Tate~\cite{dcc/ChapinT98} have experimented with computing the \BWT using a different alphabet order. This simple variant still satisfies properties \bone--\bthree, but it clearly does not bring any new theoretical insight. In the same paper, the authors also propose a variant in which rotations are sorted following a scheme inspired by reflected Gray codes. This variant shows some improvements in terms of compression, but it has never been analyzed theoretically and it does not seem to support property~\bthree.

A variant called Bijective BWT (BBWT), has been proposed in~\cite{GilScottArxiv2012} as a transformation which is bijective even without assuming that the input string~$s$ be primitive. In this variant, the output consists of the
last characters of the lexicographically sorted {\em cyclic} rotations of all factors of the Lyndon Factorization~\cite{ChenFoxLyndon1958} of~$s$. This variant has been recently shown in~\cite{cpm/BannaiKKP21} to satisfy \bone, but being based on the cyclic rotations of the Lyndon factors property \btwo\ has not been studied. As for property~\bthree, searching a pattern $P$ directly on the (compressed) transformed string takes $\Oh(|P|\log|P|)$ time~\cite{cpm/BannaiKKP19}. Note that the related variant Extended BWT~\cite{MantaciRRS07}, which takes as input a collection of strings, supports search in linear $\Oh(|P|)$ time for the problem of circular pattern matching~\cite{tcs/Egidi2020,talg/FerraginaV10,isaac/HonLST11,cpm/HonKLST12,BCLRS21_ebwt,BoucherCL0S21}.

More recently, some authors have proposed variants in which the lexicographic order is replaced by a different order relation. The interested reader can find relevant work in a recent review~\cite{DAYKIN2017}. It turns out that these variants satisfy property \bone\ in part but nothing is known with respect to properties \btwo\ and \bthree\ (an outline of a substring matching algorithm is given in \cite[Sec.~6]{tcs/DaykinMS21} without any time analysis, but is based on substrings rather than single symbols). A major problem when using ordering substantially different from the lexicographic order is that the rotations prefixed by the same substring are not necessarily consecutive in the sorted matrix. For instance, if the cyclic rotations of Figure~\ref{fig:ABWTx} are sorted according to the V-order~\cite{DaykinIliopoulosSmyth_1994}, the two rotations prefixed by $ab$ are not consecutive in the ordering. Since all the BWT-based searching algorithms work by keeping track of the rows prefixed by larger and larger pattern substrings, the fact that these rows are not consecutive makes the design of an efficient algorithm extremely difficult.

To the best of our knowledge, the only non trivial \BWT variant that fully satisfies properties \bone--\bthree\ is the {Alternating \BWT} (\ABWT). This transformation has been derived in~\cite{GesselRestivoReutenauer2012} starting from a result in combinatorics of words~\cite{CDP2005} characterizing the \BWT as the inverse of a known bijection between words and multisets of primitive necklaces~\cite{GeRe}. The \ABWT is defined as the \BWT except that when sorting rotation instead of the standard lexicographic order we use a different lexicographic order, called the {\em alternating} lexicographic order.  In the alternating lexicographic order, the first character of each rotation is sorted according to the standard order of $\Alpha$ (i.e., $a  < b < c$). However, if two rotations start with the same character we compare their second characters using the reverse ordering (i.e., $c<b<a$) and so on, alternating the standard and reverse orderings in odd and even positions. Figure \ref{fig:ABWTx} (right) shows how the rotations of an input string are sorted using the alternating ordering and the resulting \ABWT.

{The algorithmic properties of the \BWT and \ABWT are compared in~\cite{dlt/GiancarloMRRS18,abwt_jou}. It is shown that they can be both computed and inverted in linear time and that their main difference is in the definition of the LF-map, i.e. the correspondence between the characters in the first and last column of the sorted rotations matrix. In the original \BWT, the $i$-th occurrence of a character $c$ in the first column $F$ corresponds to the $i$-th occurrence of $c$ in the last column $L$, i.e., equal characters appear in the same relative order in $F$ and $L$. Instead, in the \ABWT, equal characters appear in the {\em reverse} order in $F$ and $L$. That is, the $i$-th occurrence of $c$ from the {\em top} in $F$ corresponds to the $i$-th occurrence of $c$ from the {\em bottom} in $L$. Since this modified LF-map can still be computed efficiently using \rank operations, the \ABWT can replace the \BWT for the construction of self-indices. The experiments in~\cite{dlt/GiancarloMRRS18} show that the \ABWT has essentially the same compression performance of the \BWT. However, we are not aware of any experiment using the \ABWT for indexing purposes. 

Note that in~\cite{dlt/GiancarloMRRS18, abwt_jou} the \ABWT has been studied within a larger class of transformations in which the alphabet ordering depends on the position of the characters within any cyclic rotation. Although the ``compression boosting''  property holds for all transformations in this class, in~\cite{abwt_jou} it is shown that the algorithmic techniques, based on the computation of the \rank function on the transformed string, that allow us to invert \BWT and  \ABWT in linear time cannot be applied to any other transformation in that class~\cite[Theorem~5.9]{abwt_jou}. Indeed, apart from the \BWT and \ABWT, for all the transformations studied in~\cite{abwt_jou}, the best inversion algorithm takes cubic time~\cite[Theorem~4.4]{abwt_jou}. In this paper we improve this result by providing a quadratic time algorithm for a general class of string transformations that includes the one studied in~\cite{dlt/GiancarloMRRS18,abwt_jou} (see Section~\ref{sec:lcplen}).}

\subsection{Number of runs minimization}\label{sec:runs:bkg}

The problem of minimizing the number of runs in a transformed string has been studied initially for collections of (relatively short) strings because of the relevance of this setting for bioinformatics applications~\cite{CoxBauerJakobiRosone2012,Li2014ropebwt}.
String collections are usually transformed with the Extended BWT~\cite{MantaciRRS07} (EBWT) or with an EBWT variant in which a distinct end-marker is appended to each string, making the collection ordered~\cite{BauerCoxRosoneCPM11,BauerCoxRosoneTCS2013}.
In~\cite{CoxBauerJakobiRosone2012} the authors introduce some heuristics for reordering the strings on-the-fly during the EBWT construction in order to reduce the number of runs. Experiments show that these heuristics yields a significant improvement in the overall compression. Recently, the authors of~\cite{BentleyGT19,esa/BentleyGT20} extended these results obtaining an offline linear time algorithm for finding the string reordering yielding the minimum number of runs, and showing that the optimal reordering can reduce the number of runs by a factor $\Omega(\log_{\sigma} n)$, where $n$ is the sum of the string lengths and $\sigma$ is the alphabet size.
The authors in~\cite{CGLR_DCC2023} present the first tool that guarantees to output a BWT of a string collection with minimal number of runs, in terms of reordering of input strings.
The authors of~\cite{cazauxRivals2019} consider instead the case in which the {\em same} end-marker is appended to each string and for this setting provide an $\Oh(\sigma n)$ time algorithm to find the string reordering that minimize the number of BWT runs. 
See~\cite{CenzatoL22} for a recent review of BWT variants for string collections and their properties with respect to the number of runs.

The problem of minimizing the number of runs in a single-string BWT has been studied in~\cite{esa/BentleyGT20} where the authors prove that the decision problem of finding the alphabet permutation that minimizes the number of runs in the BWT is NP-complete and the optimization variant is APX-hard. Note that, by introducing new BWT variants, we are adding a new dimension to the run minimization problem: in addition to minimizing over alphabet or string reorderings, we can also minimize over a given class of BWT variants.

\section{BWTs based on Context Adaptive Alphabet Orderings}\label{sec:genBWT}

In this section we introduce a class of string transformations that generalize the \BWT in a very natural way. Given a primitive string $s$, as in the original \BWT definition, we consider the matrix containing all its cyclic rotations. 
In the original \BWT, the matrix rows are sorted according to the standard lexicographic order. 
We generalize this concept by sorting the rows using an ordering that {\it depends on their common context}, i.e., their longest common prefix. Formally, for each string $x$ that prefixes two or more rows, we assume that an ordering $\pi_x$ is defined on the symbols of $\Alpha$. When comparing two rows which are both prefixed by $x$, their relative rank is determined by the ordering $\pi_x$. Once the matrix rows have been ordered with this procedure, the output of the transformation is the last column of the matrix, as in the original \BWT. 
Thus, these \BWT variants are based on {\em context adaptive alphabet orderings}. 
For simplicity, in the following, we call them {\em context adaptive \BWTs}.

\begin{figure}[tb]
{
{\small
$$
\begin{array}{cc@{\ \ }c@{\ \ }c@{\ \ }c@{\ \ }c@{\ \ }c@{\ \ }c@{\ \ }c@{\ 
\ }c@{\ \ }c@{\ \ }}
 & F & & & &          &  &      &    & L & \\
 &\downarrow& & & &&&& &\downarrow &\\
1 & b & a & a & a & b & a & c & a & a & \\ 
2 & b & a & c & a & a & b & a & a & a &\\ 
3 & a & c & a & a & b & a & a & a & b & \\
4  & \cellcolor[gray]{.85}a & \cellcolor[gray]{.85}a & \cellcolor[gray]{.85}b & \cellcolor[gray]{.85}a &\cellcolor[gray]{.85} a & \cellcolor[gray]{.85}a & \cellcolor[gray]{.85}b & \cellcolor[gray]{.85}a & \cellcolor[gray]{.85}c & \leftarrow s \\  
5 & \cellcolor[gray]{.85}a & \cellcolor[gray]{.85}a & \cellcolor[gray]{.85}b & \cellcolor[gray]{.85}a & \cellcolor[gray]{.85}c & \cellcolor[gray]{.85}a & \cellcolor[gray]{.85}a & \cellcolor[gray]{.85}b &  \cellcolor[gray]{.85}a & \\  
6 & \cellcolor[gray]{.85}a & \cellcolor[gray]{.85}a & \cellcolor[gray]{.85}a & \cellcolor[gray]{.85}b & \cellcolor[gray]{.85}a & \cellcolor[gray]{.85}c & \cellcolor[gray]{.85}a & \cellcolor[gray]{.85}a & \cellcolor[gray]{.85}b & \\ 
7 & a & b & a & a & a & b & a & c & a & \\ 
8 & a & b & a & c & a & a & b & a & a & \\
9 & c & a & a & b & a & a & a & b & a & \\ 
\end{array}
$$
}
}
\caption{The generalized \BWT matrix for the string $s=aabaaabac$ computed using the orderings $\pi_\epsilon = (b,a,c)$, $\pi_a = (c,a,b)$, $\pi_{aa} = (b,a,c)$, $\pi_{aaba} = (a,c,b)$, and $\pi_x = (a,b,c)$ for every other substring $x$. The horizontal arrow marks the position of the original string $s$; the last column $L$ is the output of the transformation. The range $\Rng{aa}=[4,3]$ of the three rows prefixed by $aa$ are highlighted in gray.}\label{fig:BWTx}
\end{figure}

We denote by $\Mx(s)$ the matrix obtained using this generalized sorting procedure, and by $L=\BWTx(s)$ the last column of $\Mx(s)$. Clearly $L$ depends on $s$ and the ordering used for each common prefix. Since we can arbitrarily choose an alphabet ordering for any substring $x$ of $s$, and there are $\asize!$ orderings to choose from, our definition includes a very large number of string transformations.

\begin{example}
In Figure~\ref{fig:BWTx}, the generalized \BWT matrix for the string $s=aabaaabac$ is shown. The ordering associated to the empty string $\epsilon$ is $\pi_\epsilon = (b,a,c)$ so, among the rows that have no common prefix, first we have those starting with $b$, then those starting with $a$, and finally the one starting with $c$. Since $\pi_a = (c,a,b)$, among the rows which have $a$ as their common prefix, first we have the ones 
in which $c$ is the next letter, then the ones in which $a$ is the next letter, 
followed by the ones in which $b$ is the next letter.
The complete ordering of the rows is established in a similar way on the basis of the orderings $\pi_x$. In particular, the ordering between any pair of rows is determined by $\pi_x$ where $x$ is their longest common prefix.
\end{example}

\begin{figure}
\begin{minipage}[b]{0.45\linewidth}
\centering
\includegraphics[scale=0.65]{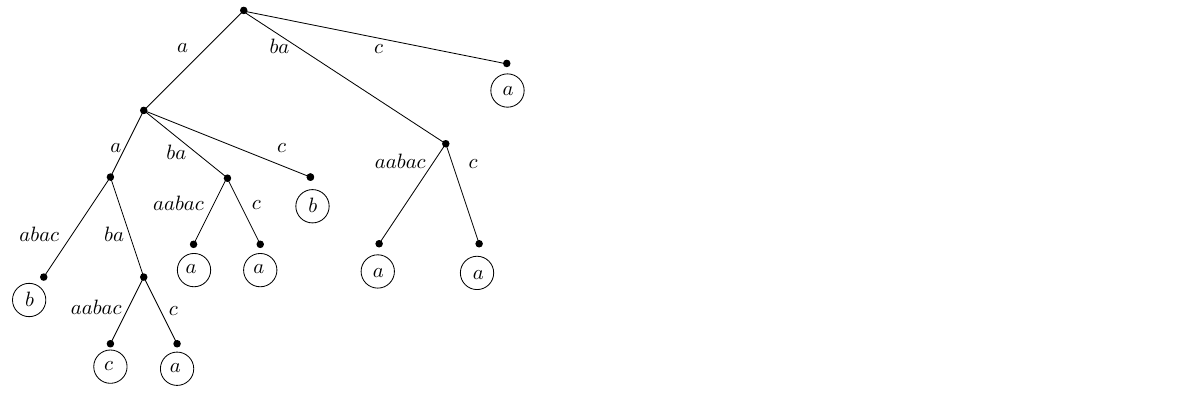}
\end{minipage}
\hfill
\begin{minipage}[b]{0.50\linewidth}
\centering
\includegraphics[scale=0.65]{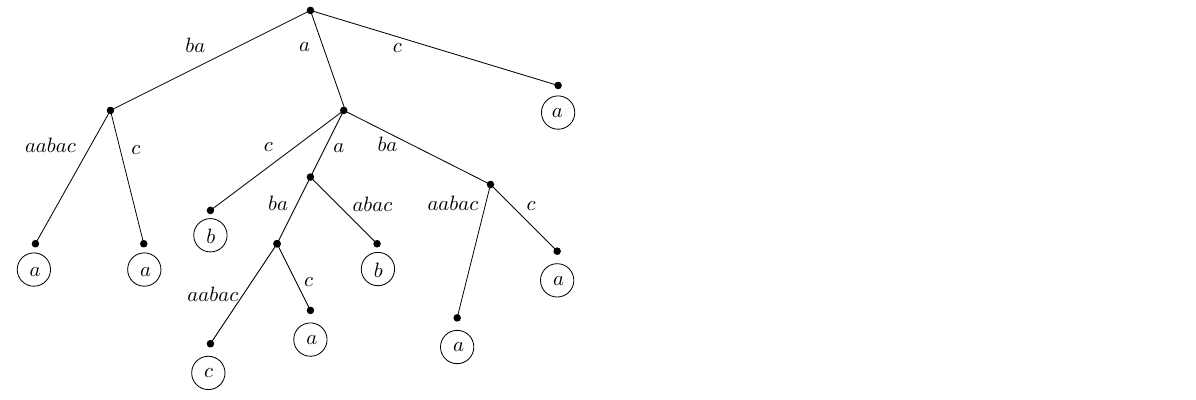}
\end{minipage}

\caption{Standard suffix tree for $s=aabaaabac$ with the symbol $c$ used as a string terminator (left), and suffix tree with edges reordered using the same orderings of Figure~\protect{\ref{fig:BWTx}} (right). To each leaf it is associated the symbol preceding in $s$ the suffix spelled by that leaf. Note that reading left to right the symbols associated to each leaf gives $\bwt(s)$ (left) and $\BWTx(s)$ (right).\label{fig:suffixTree}}
\end{figure}

This class of transformations has been mentioned in~\cite[Sect.~5.2]{FGMS2005} under the name of {\it string permutations realized by a Suffix Tree} (the definition in~\protect{\cite{FGMS2005}} is slightly more general; for example it includes the bounded context \BWT, which is not included in our class). Indeed, one can easily see that  $L=\BWTx(s)$ can also be obtained by visiting the suffix tree of~$s$ in depth first order, except that when we reach a node $u$ (including the root), we sort its outgoing edges according to their first characters using the permutation associated to the string $u_x$ labeling the path from the root to $u$. During such a visit, each time we reach a leaf, we write the symbol associated to it: the resulting string is exactly $L=\BWTx(s)$ (see Figure~\ref{fig:suffixTree} (right)).

Although in~\cite{FGMS2005} the authors could not prove the invertibility of context adaptive transformations, which we do in Section~\ref{subsec:inversion}, they observed that their relationship with the suffix tree has two important consequences: 1) they can be computed in $\Oh(n\log\sigma)$ time, and 2) they provably produce {\em highly compressible} strings, i.e., they have the ``boosting'' property of transforming a zeroth order compressor into a $k$-th order compressor.

\ignore{
To see this observe that $L=\BWTx(s)$ can be obtained visiting the suffix tree in depth first order except that when we reach a node $u$ (including the root), we sort its outgoing edges according to their first characters using the permutation associated to the string $u_x$ labeling the path from the root to $u$. During such visit, each time we reach a leaf  we write the symbol associated to it: the resulting string is exactly $L=\BWTx(s)$. 
To see that generalized \BWTs have the boosting property we observe that the proof for the \BWT (Theorem~5.4 in~\cite{FGMS2005}) is based on structural properties of the suffix tree, and can be repeated verbatim for the generalized \BWTs.}

Summing up, context adaptive transformations generalize the \BWT in two important aspects: efficient computation (linear time in $n$) and compressibility. In~\cite{FGMS2005}, the only known instances of {\em reversible} suffix tree induced transformations were the original \BWT and the bounded context \BWT. In the following, we prove that {\em all} context adaptive \BWTs defined above are invertible. Interestingly, to prove invertibility we first establish another important property of these transformations, namely that they can be used to count the number of occurrences of a pattern in $s$, which is another fundamental property of the original \BWT.

We conclude this section by observing that both the \BWT and \ABWT belong to the class we have just defined. To get the \BWT, we trivially define $\pi_x$ to be the standard $\Alpha$ ordering for every $x$, and to get the \ABWT, we define $\pi_x$ to be the standard $\Alpha$ ordering for every $x$ with $|x|$ even, and the reverse ordering for $\Alpha$ for every $x$ with $|x|$ odd.

\subsection{Counting occurrences of patterns in Context Adaptive BWTs}\label{subsec:counting}

Let $L=\BWTx(s)$ denote a context adaptive \BWT. In the following, we assume that $L$ is enriched with data structures supporting constant time rank queries as in Theorem~\ref{theo:rsa}. In this section we show that, given $L$ and the set of alphabet permutations used to build $\Mx(s)$, then  we can determine in $\Oh(\asize|x|^2)$ time the set of $\Mx(s)$ rows prefixed by $x$, for any string $x$. We preliminarily observe that, by construction, this set of rows, if non-empty, form a contiguous range inside $\Mx(s)$. This observation justifies the following definitions. 

\begin{definition}\label{def:range}
Given a string $x$, we denote by $\Rng{x} = [b_x,\ell_x]$ the range of rows of $\Mx(s)$ prefixed by $x$. More precisely, if $\Rng{x} = [b_x, \ell_x]$, then row $i$ is prefixed by $x$ if and only if it is $b_x \leq i < b_x + \ell_x$.  If no rows are prefixed $x$, we set $\Rng{x} = [0,0]$. Note that $\ell_x$ is the number of occurrences of $x$ in the circular string $s$.
\end{definition}

For technical reasons, given $x$, we are also interested in the set of rows prefixed by the strings $xc$ as $c$ varies in $\Alpha$. Clearly, these sets of rows are consecutive in $\Mx(s)$ and their union coincides with $\Rng{x}$. 

\begin{definition}\label{def:rangex}
Given a string $x$, we denote by $\Rngx{x}$ the set of $\asize+1$ integers $[b_x,\ell_1, \ell_2, \ldots, \ell_\asize]$ such that $b_x$ is the lower extreme of $\Rng{x}$ and, for $i=1,\ldots,\asize$, $\ell_i$ is the number of rows of $\Mx(s)$ prefixed by $xc_i$.
\end{definition}

Since $\Rng{x}$ is the union of the ranges $\Rng{x\xchar}$, for $\xchar\in\Alpha$, we have that, if $\Rngx{x} = [b_x,\ell_1, \ell_2, \ldots, \ell_\asize]$, then $\Rng{x}=[b_x,\sum_i \ell_i]$. Note also that the ordering of the ranges $\Rng{x\xchar}$ within $\Rng{x}$ is determined by the permutation $\pi_x$. 
As observed in Section~\ref{sec:prel}, we can assume that $L$ supports constant time rank queries. This implies that, in constant time, we are also able to count the number of occurrences of a symbol $c$ inside a substring $L[i,j]$.

\begin{example}
Let us consider the string $s=aabaaabab$ and the generalized BWT matrix illustrated in Figure \ref{fig:BWTx}.  
Since $\pi_{aa} = (b,a,c)$, we have that $\Rngx{aa}=[4,2,1,0]$,
therefore $\Rng{aa}=[4,2+1+0]=[4,3]$.
\end{example}

\begin{lemma}\label{lemma:rangex+sigma}
Given $\Rngx{x}$ and the permutation $\pi_x$, the set of values $\Rng{x\xchar_i}$ for all $\xchar_i\in\Alpha$ can be computed  in $\Oh(\asize)$ time.
\end{lemma}

\begin{proof}
If $\Rngx{x} = [b_x,\ell_1, \ell_2, \ldots, \ell_\asize]$, then
$\Rng{x\xchar_i} = [b,\ell]$ with
\begin{equation}\label{eq:j<i}
b = b_x + \sum_{j : c_j <_{\pi_x} c_i} \ell_j, \qquad
\ell = \ell_i
\end{equation}
where the summation in~\eqref{eq:j<i} is done over all $j\in\{1,2,\ldots,\asize\}$ such that $\xchar_j$ is smaller than $\xchar_i$, according to the permutation $\pi_x$.
\end{proof}

\begin{lemma}\label{lemma:newton}
Let $x=x_1 x_2 \cdots x_m$ be any length-$m$ string with $m>1$. Then, given $\Rngx{x_1 \cdots x_{m-1}}$ and $\Rngx{x_2 \cdots x_m}$, the set of values $\Rngx{x_1 \cdots x_m}$ can be computed in $\Oh(\asize)$ time.
\end{lemma}

\begin{proof}
By Lemma~\ref{lemma:rangex+sigma}, given $\Rngx{x_1 \cdots x_{m-1}}$ and $x_m$, we can compute $\Rng{x_1 \cdots x_m} = [b_x,\ell_x]$. 
In order to compute $\Rngx{x_1 \cdots x_m}$, we additionally need the number of rows prefixed by $x_1 x_2 \cdots x_m \xchar$, for any $\xchar\in\Alpha$. These numbers can be obtained by first computing the ranges $\Rng{x_2 \cdots x_m \xchar}$ using again Lemma~\ref{lemma:rangex+sigma}. The number of rows prefixed by $x_1 x_2 \cdots x_m \xchar$ can be obtained by counting the number of $x_1$ in the portions of $L$ corresponding to each range $\Rng{x_2 \cdots x_m \xchar}$. The counting takes $\Oh(\asize)$ time since we are assuming $L$ supports constant time \rank as in Theorem~\ref{theo:rsa}. 
\end{proof}

\begin{theorem}\label{theorem:rangex}
Suppose we are given $\BWTx(s)$ with constant time rank support, and the set of permutations used to compute the matrix $\Mx(s)$. Then, given any string $x=x_1 x_2 \cdots x_p$, the range of rows $\Rng{x}$ prefixed by $x$ can be computed in $\Oh(\asize p^2)$ time and $\Oh(\asize p)$ space. 
\end{theorem}

\begin{proof}
We need to compute $\Rng{x_1 x_2 \cdots x_p}$. To this end we consider the following scheme, inspired by the Newton finite difference formula:
$$
\begin{array}{clllll}
\Rngx{x_1} & \Rngx{x_1 x_2} & \Rngx{x_1 x_2 x_3} & \cdots & \Rngx{x_1 x_2 \cdots x_{p-1}} & \Rngx{x_1 x_2 \cdots x_p} \\
\Rngx{x_2} & \Rngx{x_2 x_3} & \Rngx{x_2 x_3 x_4} & \cdots &  \Rngx{x_2 \cdots x_{p}} & \\
\Rngx{x_3} & \Rngx{x_3 x_4} &\multicolumn{1}{c}{ \cdots} & \\
\vdots \\
\Rngx{x_p}
\end{array}
$$
Using Lemma~\ref{lemma:newton}, we can compute $\Rngx{x_i \cdots x_j}$ given $\Rngx{x_i \cdots x_{j-1}}$ and $\Rngx{x_{i+1} \cdots x_j}$. Thus, from two consecutive entries in the same column, we can compute one entry in the following column.  
To compute $\Rng{x_1 x_2 \cdots x_p}$ we can for example perform the computation 
bottom-up, proceeding row by row. In this case, we are essentially computing the ranges corresponding to $x_p$, $x_{p-1} x_p$, $x_{p-2} x_{p-1} x_p$ and so on, in a sort of backward search. However, we can also perform the computation top down, diagonal by diagonal, and in this case, we are computing the ranges corresponding to $x_1$, $x_1 x_2$, and so on, up to $x_1 \cdots x_p$. In both cases, the information one needs to store from one iteration to the next is $\Oh(p)$ $\Rngx{\cdot}$ values, which take $\Oh(\asize p)$ words. By Lemma~\ref{lemma:newton}, the computation of each value takes $\Oh(\asize)$ time, so the overall complexity is $\Oh(\asize p^2)$ time.  
\end{proof}

\begin{example}
For the transformation described in Figure~\ref{fig:BWTx}, the computation of Theorem~\ref{theorem:rangex} for computing $\Rng{aba}$ works as follows 
$$
\begin{array}{cll}
\Rngx{a}=[3,1,3,2] & \quad\Rngx{ab}=[7,2,0,0] & \quad\Rngx{aba}=[7,1,0,1] \\
\Rngx{b}=[1,2,0,0] & \quad\Rngx{ba}=[1,1,0,1] \\
\Rngx{a}=[3,1,3,2] 
\end{array}
$$
Finally, since $\Rngx{aba} = [7,1,0,1]$, we have $\Rng{aba}=[7,1+0+1] = [7,2]$.
\end{example}

Note that our scheme for the computation of $\Rng{x}$ is based on the computation of $\Rngx{y}$ for $\Oh(p^2)$ substrings $y$ of $x$. If $x$ has many repetitions, the overall cost could be less than quadratic. In the extreme case, $x=a^p$, $\Rng{x}$ can be computed in $\Oh(\asize p)$ time.

\subsection{Inverting Context Adaptive BWTs}\label{subsec:inversion}

We now show that the machinery we set up for counting occurrences can be used to retrieve $s$ given $\BWTx(s)$, thus to invert any context adaptive \BWT. 

\begin{lemma}\label{lemma:nextchar}
Given $\Rngx{x} = [b_x,\ell_1,\ell_2,\ldots,\ell_\asize]$ and a row index $i$ with $b_x \leq i < b_x + \sum_{j=1}^\asize \ell_j$,  the $(|x|+1)$-st character of row $i$ can be computed in $\Oh(\asize)$ time.
\end{lemma}

\begin{proof}
Let $\rho_{1}, \ldots, \rho_{\asize}$ denote the alphabet symbols reordered according to permutation~$\pi_x$, and let $\ell'_1, \ldots, \ell'_\asize$ denote the values $\ell_1,\ldots,\ell_\asize$ reordered according to the same permutation. Since $i\in\Rng{x}$, row $i$ is prefixed by $x$. Since the rows prefixed by~$x$ are sorted in their $(|x|+1)$-st position according to $\pi_x$, the $(|x|+1)$-st symbol of row $i$ is the symbol $\rho_j$ such that 
$$ 
b_x + \sum_{1 \leq h<j} \ell'_{h} \;\leq\; i \;<\;  
b_x + \sum_{1 \leq h\leq j} \ell'_{h}
$$
\end{proof}

\begin{theorem}\label{theo:inversion}
Given $\BWTx(s)$ with constant time rank support, the permutations $\pi_x$ used to build the matrix $\Mx(s)$, and the row index $I$ containing $s$ in $\Mx(s)$, the original string $s$ can be recovered in $\Oh(\asize |s|^2)$ time and $\Oh(\asize |s|)$ working space.
\end{theorem}

\begin{proof}
Let $s = s_1 s_2 \cdots s_n$. From $\BWTx(s)$, in $\Oh(n)$ time we retrieve the number of occurrences of each character in $s$ and hence the ranges $\Rng{\xchar_1}$, $\Rng{\xchar_2}$, \ldots, $\Rng{\xchar_\asize}$. From those and the row index $I$, we retrieve the first character of $s$, i.e. $s_1$. Next, counting the number of occurrences of $s_1$ in the ranges of $\BWTx(s)$ corresponding to $\Rng{\xchar_1}$, $\Rng{\xchar_2}$, \ldots, $\Rng{\xchar_\asize}$, we compute $\Rngx{s_1}$.
Finally, we show by induction that, for $m=1,\ldots,n-1$, given $\Rngx{s_1 s_2 \cdots s_m}$, we can retrieve $s_{m+1}$ and $\Rngx{s_1 s_2 \cdots s_{m+1}}$ in $\Oh(m\asize)$ time. By using Lemma~\ref{lemma:nextchar}, from $\Rngx{s_1 s_2 \cdots s_m}$ and $i$ we retrieve $s_{m+1}$. Next, assuming we maintained the ranges 
$\Rngx{s_j \cdots s_m}$, for $j=1,\ldots,m$ we can compute $\Rngx{s_j \cdots s_{m+1}}$ adding one diagonal to the scheme shown in the proof of Theorem~\ref{theorem:rangex}. By Lemma~\ref{lemma:newton}, the overall cost is $\Oh(\asize |s|^2)$ time as claimed. 
\end{proof}

The above theorem establishes that all context adaptive \BWTs are invertible. Note that in our definition, the alphabet ordering $\pi_x$ associated to $x$ can depend on the whole string $x$; in this sense the context has full memory. We consider this an important conceptual result. However, from a practical point of view, a transformation whose definition requires the specification of $\Oh(|s|)$ alphabet permutations appears rather cumbersome. For this reason, in the following, we consider some subclasses of transformations in which the permutations associated to each prefix are defined by ``simple'' rules. We show that, in some cases, nontrivial generalized \BWTs have simpler and more efficient inversion algorithms.

\subsection{Special case: ordering based on context length}\label{sec:lcplen}

In~\cite{dlt/GiancarloMRRS18}, the authors introduced a set of generalized transformations that turns out to be a subclass of context adaptive \BWTs. Given a $k$-tuple  of alphabet permutations $K = (\pi_0, \pi_1, \ldots, \pi_{k-1})$, using the notation of this paper, the transformation $\BWT_K$ in~\cite{dlt/GiancarloMRRS18} is defined as the context adaptive \BWT in which the permutation $\pi_x$ associated to the string $x$ is $\pi_\ell$ where $\ell = |x| \bmod k$. Hence, the permutation associated to each string only depends on its depth, and the $k$-tuple $K = (\pi_0, \pi_1, \ldots, \pi_{k-1})$ completely determines the transformation. In~\cite{dlt/GiancarloMRRS18,abwt_jou}, it was shown that for every $K$ the transformation $\BWT_K$ can be inverted in $\Oh(|s|^3)$ time; Theorem~\ref{theo:inversion} therefore constitutes an alternative faster inversion algorithm. To our knowledge, no faster algorithm is known, even for $k=2$, when the whole transformation depends on just two alphabet permutations.

\begin{example}
Let us consider $k=3$ and the $k$-tuple of alphabet permutations $K=(\pi_0,\pi_1,\pi_2)$, where $\pi_0=(c,a,b)$, $\pi_1=(b,c,a)$ and $\pi_2=(b,a,c)$. The $\BWT_K$ matrix for the string $s=aabaaabac$ is shown in Figure \ref{fig:BWTk}. If $x$ is the longest common prefix between two rows, their ordering depend on the respective characters at the $(|x|+1)$-th position, according to the permutation $\pi_{|x|\bmod 3}$. For instance, $abacaabaa<abaaabaca$, since $aba$ is the longest common prefix and $c<a$, according to $\pi_0$.
\end{example}

\begin{figure}[tb]
{
{\small
$$
\begin{array}{cc@{\ \ }c@{\ \ }c@{\ \ }c@{\ \ }c@{\ \ }c@{\ \ }c@{\ \ }c@{\ \ }c@{\ \ }l@{\ \ }}
 & F & & & &          &  &      &    & L & \\
 &\downarrow& & & &&&& &\downarrow    &   \\
1 & c & a & a & b & a & a & a & b & a  &   \\ 
2 & a & b & a & c & a & a & b & a & a  &   \\
3 & a & b & a & a & a & b & a & c & a  &    \\ 
4 & a & c & a & a & b & a & a & a & b  &    \\
5  & a & a & b & a & c & a & a & b & a &  \leftarrow s  \\  
6 & a & a & b & a & a & a & b & a & c  &    \\  
7 & a & a & a & b & a & c & a & a & b  &    \\ 
8 & b & a & a & a & b & a & c & a & a  &    \\ 
9 & b & a & c & a & a & b & a & a & a  &    \\ 
\end{array}
$$
}
\caption{The $\BWT_K$ matrix for the string $s=aabaaabac$ computed using the triple of alphabet permutations $K=\{(c,a,b),(b,c,a),(b,a,c)\}$. The horizontal arrow marks the position of the original string $s$; the last column $L$ is the output of the transformation.}\label{fig:BWTk}
}
\end{figure}

\subsection{Special case: $\pm$ ordering}\label{sec:pm}

Given a permutation $\pi$ of the alphabet $\Alpha$, we denote by $\pi^R$ the reversal of $\pi$, that is, the permutation such that, for each pair of symbols $\xchar_i$, $\xchar_j$ in $\Alpha$,
$$
\pi^R(\xchar_i) < \pi^R(\xchar_j) \quad\Longleftrightarrow\quad 
\pi(\xchar_i) > \pi(\xchar_j).
$$


Let $\pi$ denote an arbitrary permutation of $\Alpha$. We consider the subclass of context adaptive transformations in which the permutation $\pi_x$ associated to each substring $x$ can be either $\pi$ or its reversal $\pi^R$. Once $\pi$ is established, for each string $x$ we only need an additional bit to specify the ordering $\pi_x$. For this subclass we say that the matrix $\Mx(s)$ is based on a {\em $\pm$~ordering} (see example in Figure~\ref{fig:BWTrev_ord}).  In this section, we show that, for the transformations in this subclass, the space and time for inversion can be reduced by a factor~$\asize$. 

\begin{figure}[tb]
{\small
$$
\begin{array}{cc@{\ \ }c@{\ \ }c@{\ \ }c@{\ \ }c@{\ \ }c@{\ \ }c@{\ \ }c@{\ \ }c@{\ \ }l@{\ \ }}
 & F              &      &    & &          &  &      &    & L &\\
                  &\downarrow & & & &&&& &\downarrow  &\\
1               & b  & a & a & a & b & a & c & a & a &\\ 
2& \cellcolor{brown!20}b & \cellcolor{brown!20}a & \cellcolor{brown!20}c & \cellcolor{brown!20}a & \cellcolor{brown!20}a & \cellcolor{brown!20}b & \cellcolor{brown!20}a & \cellcolor{brown!20}a & \cellcolor{brown!20}a & \\ 
{3}               & a  & c & a & a & b & a & a & a & b & \\
{4}               & a  & a & b & a & c & a & a & b & a &\\  
{5}               & a  & a & b & a & a & a & b & a & c &  \leftarrow s\\  
{6}               & a  & a & a & b & a & c & a & a & b &\\ 
{7}             & \cellcolor{cyan!40}a & \cellcolor{cyan!40}b & \cellcolor{cyan!40}a & \cellcolor{cyan!40}a & \cellcolor{cyan!40}a & \cellcolor{cyan!40}b & \cellcolor{cyan!40}a & \cellcolor{cyan!40}c & \cellcolor{cyan!40}a &\\ 
{8}               & \cellcolor{cyan!40}a & \cellcolor{cyan!40}b & \cellcolor{cyan!40}a & \cellcolor{cyan!40}c & \cellcolor{cyan!40}a & \cellcolor{cyan!40}a & \cellcolor{cyan!40}b & \cellcolor{cyan!40}a & \cellcolor{cyan!40}a & \\
{9}               & c  & a & a & b & a & a & a & b & a & \\ 
\end{array}
$$
}
\caption{Example of a transformation based on a $\pm$~ordering. Let $\pi=(b,a,c)$ so that $\pi^R=(c,a,b)$.
The generalized \BWT matrix $\Mx(s)$ for the string $s=aabaaabac$ is computed using the following permutations
$\pi_{\epsilon}=\pi$, $\pi_a = \pi^R$, $\pi_{aa} = \pi$, $\pi_{aaba} = \pi^R$, and $\pi_x = \pi$ for every other substring $x$. The horizontal arrow marks the position of the original string $s$; the last column $L$ is the output of the transformation. The range $\Rng{aba}=[7,2]$ of the rows prefixed by $aba$ are coloured in cyan. In light brown, the row of the range $\Rng{bac}=[2,1]$.}
\label{fig:BWTrev_ord}
\end{figure}

\ignore{
\begin{example}
Let us consider the string $s=aabaaabac$ and the permutation $\pi=(b,a,c)$ so that $\pi^R=(c,a,b)$. In Figure \ref{fig:BWTrev_ord},  the generalized \BWT matrix $\Mx(s)$ based on a $\pm$ ordering is shown. In particular, $\Mx(s)$ is computed by using $\pi_{\epsilon}=\pi$, $\pi_a = \pi^R$, $\pi_{aa} = \pi$, $\pi_{aaba} = \pi^R$, and $\pi_x = \pi$ for every other substring $x$. 
\end{example}
}

\begin{lemma}\label{lemma:newton+-}
Let $\Mx(w)$ be based on a $\pm$~ordering, and let
$x=x_1 x_2 \cdots x_m$ be any length-$m$ string, with $m>1$. 
Then, given $\Rng{x_1 \cdots x_{m-1}}$, $\Rng{x_2 \cdots x_m}$ and $\Rng{x_2\cdots x_{m-1}}$, we can compute $\Rng{x_1 \cdots x_m}$ in $\Oh(1)$ time.
\end{lemma}

\begin{proof}
Let $y=x_1 \cdots x_{m-1}$ and $z=x_2 \cdots x_{m-1}$. Recall that there is a bijection between the rows in $\Rng{y}$ and the rows in $\Rng{z}$ whose last symbol is $x_1$. We exploit this bijection to find $\Rng{y x_m}$ given $\Rng{z x_m}$. The size of $\Rng{y x_m}$ is equal to the number of rows in $\Rng{z x_m}$ ending with $x_1$, so to completely determine $\Rng{y x_m}$ we just need to compute how many rows in $\Rng{y}$ {precede} $\Rng{y x_m}$. If $\pi_y = \pi_z$ this number is equal to the number of rows in $\Rng{z}$ {\em above} $\Rng{z x_m}$ and ending with $x_1$. If  $\pi_y \neq \pi_z$, since necessarily $\pi_y = \pi_z^R$, this number is equal to the number of rows in $\Rng{z}$ {\em below} $\Rng{z x_m}$ and ending with $x_1$. Since counting the number of rows ending in $x_1$ in a given range can be done using rank operations on $\BWTx(s)$, the lemma follows. 
\end{proof}

\begin{example}
Consider again the example of Fig.~\ref{fig:BWTrev_ord}. Let $x=abac$, so $y=aba$ and $z=ba$. It is $\Rng{y}=[7,2]$, $\Rng{z}=[1,2]$ and $\Rng{zc}=[2,1]$. The rows of $\Rng{y}$ are highlighted in cyan, the rows of $\Rng{zc}$ are highlighted in light brown in Fig. \ref{fig:BWTrev_ord}. The number of rows in $\Rng{zc}=[2,1]$ ending with $a$ is $1$. Such a number gives the size of $\Rng{yc}=\Rng{abac}$. Moreover, since $\pi_y = \pi_z=\pi$, we need to count the number of rows in $\Rng{z}=[1,2]$ above $\Rng{zc}=[2,1]$ and ending with $a$. Such a number is $1$. This means that $1$ row in $\Rng{y}=[7,2]$ precedes $\Rng{yc}=\Rng{abac}$. Hence, $\Rng{abac}=[8,1]$. 
If we consider $x=aa$, it is $y=a$ and $z=\epsilon$, $\Rng{y}=\Rng{a}=\Rng{za}=[3,6]$, $\Rng{z}=\Rng{\epsilon}=[1,9]$. The number of rows in $\Rng{za}=[3,6]$ ending with $a$ is $3$. So, $3$ is the size of $\Rng{aa}$. Since $\pi_a \neq \pi_\epsilon$,  we have to count the number of rows in $\Rng{z}=[1,9]$ below $\Rng{za}=[3,6]$ and ending with $a$. Such a number is $1$. This means that there is $1$ row in $\Rng{y}=\Rng{a}=[3,6]$ that precedes $\Rng{ya}=\Rng{aa}$, hence $\Rng{aa}=[4,3]$.
\end{example}

\begin{lemma}\label{lemma:rangex+-}
Suppose $\Mx(s)$ is based on a $\pm$~ordering. Given $\BWTx(s)$ with constant time rank support, and any string $x=x_1 x_2 \cdots x_p$, we can compute the range of rows prefixed by $x$ in $\Oh(p^2)$ time and $\Oh(p)$ space.  
\end{lemma}

\begin{proof}
We reason as in the proof of Theorem~\ref{theorem:rangex}, except that because of Lemma~\ref{lemma:newton+-} we work with $\Rng{\cdot}$ instead of $\Rngx{\cdot}$. The thesis follows observing that each entry takes $\Oh(1)$ space and can be computed in $\Oh(1)$ time.
\end{proof}

\begin{theorem}\label{theo:inversion+-}
Suppose $\Mx(s)$ is based on a $\pm$~ordering. Given $\BWTx(s)$ with constant time rank support and the row index $i$ containing $s$ in $\Mx(s)$, we can retrieve the original string $s$ in $\Oh(|s|^2)$ time and $\Oh(|s|)$ working space.
\end{theorem}


\section{BWTs based on local orderings}\label{sec:local}

In this section, we consider the context adaptive transformations in which the alphabet ordering $\pi_x$ associated to each string $x$ only depends on the last $k$ symbols of $x$, where $k$ is fixed. In the following, we refer to these string transformations as {\em \BWTs based on local orderings}. 
We show that local ordering transformations have properties very similar to the original \BWT, since they can be inverted in linear time and also support the search of a pattern in the original text in time proportional to the pattern length.

\begin{figure}[htb]
{
{\small
$$
\begin{array}{cc@{\ \ }c@{\ \ }c@{\ \ }c@{\ \ }c@{\ \ }c@{\ \ }c@{\ \ }c@{\ \ }c@{\ \ }l@{\ \ }}
 & F & & & &          &  &      &    & L & \\
 &\downarrow& & & &&&& &\downarrow   &\\
 1 & \cellcolor{cyan!40}b & \cellcolor{cyan!40}a & a & a & b & a & c & a & a &\\ 
 2& \cellcolor{brown!20}b & \cellcolor{brown!20}a & c & a & a & b & a & a & a &\\
 3& c & a & a & b & a & a & a & b & a &\\ 
 4& a & b & a & a & a & b & a & c & a &\\ 
 5& a & b & a & c & a & a & b & a & \underline{a} & \\ 
 6& a & a & b & a & a & a & b & a & \underline{c} &  \leftarrow s\\ 
 7& a & a & b & a & c & a & a & b & \underline{a} & \\ 
 8& \cellcolor{cyan!40}a & a & a & b & a & c & a & a & \cellcolor{cyan!40}b &\\ 
 9 & \cellcolor{brown!20}a & c & a & a & b & a & a & a & \cellcolor{brown!20}\underline{b} & \\
\end{array}
$$
}
}
\caption{The local ordering \BWT matrix for the string $s=aabaaabac$ computed using the orderings $\pi_\epsilon = (b,c,a)$, $\pi_a = (b,a,c)$, $\pi_b = \pi_c = (a,b,c)$. Here, the alphabet orderings associate to each non-empty string depend only on the last symbol of the string, i.e. $k=1$. The horizontal arrow marks the position of the original string $s$; the last column $L$ is the output of the transformation. The rows starting with $ba$ (highlighted in cyan and light brown, respectively) are in a order-preserving correspondence to the rows starting with $a$ and ending with $b$. The last character of each BWT-run is underlined.}\label{fig:BWTloc}
\end{figure}
We start by analyzing the case $k=1$. For such local ordering transformations, the matrix $\Mx(s)$ depends on only $\asize+1$ alphabet orderings: one for each symbol plus the one used to sort the first column of $\Mx(s)$. See Figure~\ref{fig:BWTloc} for an example. The following lemma establishes an important property of local ordering transformations. 

\begin{lemma}\label{lemma:almostbwt}
If $\Mx(s)$ is based on a local ordering with $k=1$, then for any pair of characters $x_1$ and $x_2$,  there is an order-preserving bijection between the set of rows starting with $x_1x_2$ and the set of rows starting with $x_2$ and ending with $x_1$.
\end{lemma}

\begin{proof}
Note that both sets of rows contain a number of elements equal to the number of occurrences of $x_1 x_2$ in the circular string $s$. In the following, we write $s[i\cdots]$ to denote the cyclic rotation of $s$ starting with $s[i]$. Assume that rotations $s[i\cdots]$ and $s[j\cdots]$ both start with $x_2$ and end with $x_1$ and let $h>1$ denote the first  column in which the two rotations differ.
Rotation $s[i\cdots]$ precedes $s[j\cdots]$ in $\Mx(s)$ if and only if $s[i+h-1]$ is smaller than $s[j+h-1]$ according to the alphabet ordering associated to symbol $s[i+h-2]=s[j+h-2]$. 
The two rotations $s[i-1 \cdots]$ and $s[j-1 \cdots]$ both start with $x_1 x_2$ and their relative position also depends on the relative ranks of $s[i+h-1]$ and $s[j+h-1]$ according to the alphabet ordering associated to symbol $s[i+h-2]=s[j+h-2]$. Hence the relative order of  $s[i-1 \cdots]$ and $s[j-1 \cdots]$ is the same as the one of $s[i\cdots]$ and $s[j\cdots]$.
\end{proof}

\begin{example}
    Consider the string $s=aabaaabac$ and the local ordering \BWT computed using the orderings $\pi_\epsilon = (b,c,a)$, $\pi_a = (b,a,c)$, $\pi_b = \pi_c = (a,b,c)$ by using $k=1$. As proved in Lemma \ref{lemma:almostbwt}, the bijection between the set of rows $1$ and $2$ starting with $ba$ (highlighted in cyan and light brown, respectively) and the set of rows $8$ and $9$ starting with $a$ and ending with $b$ is order-preserving. Analogously, the bijection mapping the rows $5$, $6$, and $7$ starting with $aa$ to the rows $4$, $5$, and $7$, respectively.  
\end{example}

Armed with the above lemma, we now show that for local ordering transformations we can establish much stronger results than the ones provided in Section~\ref{subsec:counting}.

\begin{lemma}\label{lemma:newton:local}
Suppose $\BWTx(s)$ is based on a local ordering with $k=1$ and supports constant time rank queries. Let $x=x_1 x_2 \cdots x_m$ be any length-$m$ string with $m>1$. Then, given $\Rng{x_1 x_2}$, $\Rng{x_2}$ and $\Rng{x_2 \cdots x_m}$, the value $\Rng{x_1 \cdots x_m}$ can be computed in $\Oh(1)$ time.
\end{lemma}

\begin{proof}
By Lemma~\ref{lemma:almostbwt}, there is an order preserving bijection between the rows in $\Rng{x_1 x_2}$ and those in $\Rng{x_2}$ ending with $x_1$. In this bijection, the rows in $\Rng{x_1 \cdots x_m}$ correspond to those in $\Rng{x_2 \cdots x_m}$ ending with $x_1$. Because of this bijection, if, among the ordered set of rows starting with $x_2$ and ending with $x_1$, those prefixed by $x_2\cdots x_m$ are in positions $r+1, \ldots, r+h$, then, among the rows starting with $x_1 x_2$, those prefixed by $x_1 x_2 \cdots x_m$ are {\em consecutive} and in positions $r+1,\ldots,r+h$. The lemma follows since if  $\Rng{x_2} = [b,\ell]$ and $\Rng{x_2 \cdots x_m} = [b',\ell']$, we have
$$
r = \rank_{x_1}(L,b'-1) - \rank_{x_1}(L,b-1),\hfill
h =  \rank_{x_1}(L,b'+\ell'-1) - \rank_{x_1}(L,b'-1),
$$
and, if $\Rng{x_1 x_2}=[\bar{b},\bar{\ell}]$, it is $\Rng{x_1 \cdots x_m} = [\bar{b}+r,h]$. 
\end{proof}

\begin{theorem}\label{theorem:rangex:local}
Suppose $\BWTx(s)$ is based on a local ordering with $k=1$ and supports constant time rank queries. After a $\Oh(\asize^2)$ time preprocessing, given any string $x=x_1 x_2 \cdots x_p$, the range of rows prefixed by $x$ can be computed in $\Oh(p)$ time and $\Oh(\asize^2 + p)$ space. 
\end{theorem}

\begin{proof}
We reason as in the proof of Theorem~\ref{theorem:rangex}, except that because of Lemma~\ref{lemma:newton:local} we can work with $\Rng{\cdot}$ instead of $\Rngx{\cdot}$ and we only need to compute the first two columns and the diagonal. In the preprocessing step, we compute $\Rng{\xchar_i}$ and $\Rng{\xchar_i\xchar_j}$ for any pair $(\xchar_i, \xchar_j) \in \Alpha^2$. During the search phase, we compute each diagonal entry in constant time.
\end{proof}

Another immediate consequence of Lemma~\ref{lemma:almostbwt} is that we can efficiently ``move back in the text'' as in the original \BWT. Note this operation is the base for \BWT inversion and for snippet extraction and locate operations on FM-indices~\cite{Ferragina:2005}.

\begin{lemma}
Suppose $\BWTx(s)$ is based on a local ordering with $k=1$ and supports constant time rank and access queries. Then, after a $\Oh(\asize^2)$ time preprocessing, given a row index $i$ we can compute in $\Oh(1)$ time the index of the row obtained from the $i$-th row with a circular right shift by one position.
\end{lemma}

\begin{proof}
It suffices to compute the first and last symbol of row $i$ and then apply Lemma~\ref{lemma:almostbwt}.
\end{proof}

\begin{corollary}
If $\BWTx(s)$ is based on a local ordering with $k=1$ and supports constant time rank and access queries, $\BWTx(s)$ can be inverted in $\Oh(\asize^2+|s|)$ time and $\Oh(\asize^2)$ working space.
\end{corollary}

The notion of local ordering can be generalized to contexts of size $k>1$. The resulting transformations depend on $1+\sigma +\sigma^2 + \cdots + \sigma^k$ alphabet permutations, one for each string over $\Alpha$ of length at most $k$. Once these permutations have been chosen, with the notation of Section~\ref{sec:genBWT}, we consider the context adaptive transformations in which the alphabet ordering $\pi_x$ associated to each string $x$ only depends on the last $k$ symbols of $x$, or to the last $|x|$ symbols if $|x| < k$. Lemma~\ref{lemma:almostbwt} can be generalized to show that, for any $(k+1)$-tuple $x_1, \ldots, x_{k+1}$, there is an order preserving bijection between the rows prefixed by $x_2 \cdots x_{k+1}$ and ending with $x_1$ and the rows prefixed by $x_1 \cdots x_{k+1}$. As a consequence, search and inversion can still be performed in linear time with the only difference that the preprocessing phase now takes $\Oh(\asize^{k+1})$ time and space, since we need to compute and store the values $R[x]$ for all strings $x$ of length up to $k+1$.

\subsection{Local orderings and $r$-index} \label{sec:rindex}

The \rindex~\cite{GNPjacm19} is a recent variant of the FM-index, still based on the BWT, introduced to efficiently support the locate operation on highly repetitive collections.
The locate operation is the task of determining the positions, in the original input $s$, of all the occurrences of the pattern. This is usually done storing a subset of the Suffix Array entries (these are called Suffix Array samples). 
The main feature of the $r$-index is that it supports the locate operation using a number of Suffix Array samples equal to the number of runs in the BWT. If the input contains many repetitions such number is much smaller than the number of Suffix Array samples used by the standard FM-index, usually $\Theta(n/\log^c n)$ for some constant~$c>1$.

In this section we show that it is possible to implement an \rindex also on the top of a BWT based on a local ordering. We start considering the case $k=1$. The crucial ingredient of the \rindex is the so called Toehold Lemma: this result ensures that when we search for a pattern $x$, in addition to the range of rows prefixed by $x$, we also obtain the position in $s$ of at least one occurrence of $x$ (assuming such occurrence exists). 
To prove the Toehold Lemma for local orderings, we proceed as in Lemma~2 in~\cite{tcs/BannaiGI20} and we logically mark every character in $L$ which is the last character in a BWT-run, and we store the position in $s$ of the marked characters.  In addition, for each character $c$, we store the position in $s$ of the last row prefixed by~$c$.

\begin{lemma} \label{lem:rindex}
Let $x = x_1 \cdots x_p$ be a string that occurs in~$s$. Then, using a \BWT based on local ordering, in $\Oh(p)$ time we can compute, in addition to the range $\Rng{x_1 \cdots x_p}$ of rows prefixed by $x$, the position in $s$ of the last row in such a range. 
\end{lemma}

\begin{proof}
We proceed by induction on $j=p,p-1,\ldots,1$. For $j=p$ the position in $s$ of the last row prefixed by $x_p$ is obtained by $\Rng{x_p}$, which is computed in the preprocessing phase of Theorem~\ref{theorem:rangex:local}.  For $j<p$, assume that we know the range $\Rng{x_{j+1} \cdots x_p}$ and the position $\pos_{j+1}$ in $s$ of the last row in that range. 
Since there is an order preserving bijection between the rows in $\Rng{x_{j+1} \cdots x_p}$ ending with $x_j$ and the rows in $\Rng{x_{j} \cdots x_p}$, the last row in $\Rng{x_{j} \cdots x_p}$ corresponds to the last row in 
$\Rng{x_{j+1} \cdots x_p}$ ending with $x_j$. If the latter coincides with the last row in $\Rng{x_{j+1} \cdots x_p}$, i.e. $x_j$ is the last symbol in the portion of $L$ corresponding to $\Rng{x_{j+1} \cdots x_p}$,  
then the position of the last row in $\Rng{x_{j} \cdots x_p}$ in $s$ is simply $\pos_{j+1} -1$. If $x_j$ is not the last symbol in the portion of $L$ corresponding to $\Rng{x_{j+1} \cdots x_p}$, then the last $x_j$ in that range will be the last of a BWT-run and therefore will be a marked position. Its position in $s$ will be among the ones we stored, and this will coincide with the position of the last row in $\Rng{x_{j} \cdots x_p}$. 
\end{proof}

\begin{example}
Let us consider the following string $s$ and the column $L$ of its local ordering BWT-matrix, as shown in Fig. \ref{fig:BWTloc}. The column $F$ of the matrix is here also reported.

    \[
    \begin{matrix}
F= & b & \tikzmarknode{I}{b} & \tikzmarknode{L}{c} &  a         & a         & a &    a       & a & \tikzmarknode{M}{a} \\
   &  &  &  &           &           &  &           &   &  \\
s= & a & a & b & \tikzmarknode{A}{a} & \tikzmarknode{B}{a} & a & \tikzmarknode{C}{b} & \tikzmarknode{N}{a} & 
\tikzmarknode{D}{c} \\
   &  &  &  &           &           &  &            &   &    \\
 L= & a & a & a & a     & \tikzmarknode{E}{\underline{a}} & \tikzmarknode{F}{\underline{c}} & \tikzmarknode{G}{\underline{a}} &  b &  \tikzmarknode{H}{\underline{b}}   \\
    & 1 & 2 & 3 &  4         &  5         & 6 &    7       & 8 & 9 \\
    \end{matrix}
    \]

    \begin{tikzpicture}[remember picture, overlay]
        \draw[gray!80,  thick][->] (E.north) -- (B.south);    
        \draw[gray!80,  thick][->] (F.north) -- (D.south);    
        \draw[gray!80,  thick][->] (G.north) -- (A.south);    
        \draw[gray!80,  thick][->] (H.north) -- (C.south);    
        \draw[gray!80,  thick, dashed][->] (I.south) -- (C.north);    
        \draw[gray!80,  thick, dashed][->] (L.south) -- (D.north);    
        \draw[gray!80,  thick, dashed][->] (M.south) -- (N.north);    
    \end{tikzpicture}

The local ordering \BWT matrix for $s$ computed using the ordering $\pi_\epsilon = (b,c,a)$, $\pi_a = (b,a,c)$, $\pi_b = \pi_c = (a,b,c)$, as reported in Fig.~\ref{fig:BWTloc}. The last character of each BWT-run in $L$ is underlined. They are $L[5]$, $L[6]$, $L[7]$ and $L[9]$, and the corresponding positions in $s$ are $5$, $9$, $4$ and $7$, respectively. The correspondence between each underlined symbol in $L$ and its position in $L$ is represented by an arrow. In addition, for each character $c$, we store the position in $s$ of the last row prefixed by~$c$. For each of such rows, the correspondent position in $s$ if highlighted by a dashed arrow from the column $F$ to $s$. Such positions in $s$ are $7$ (row $2$), $9$ (row $3$), and $8$ (row $9$).

Let us consider the string $x = baa$.
For $j=3$, the position $p_3=8$ in $s$ of the last row prefixed by $x_3=a$ is obtained by $\Rng{a}=[4,6]$ (i.e., the row $9$).
For $j=2$, we know that there is an order preserving bijection between the rows in $\Rng{a}=[4,6]$ ending with $a$ and the rows in $\Rng{aa}=[6,3]$. Since $x_2=a$ \emph{is not} the last symbol in the portion of $L$ corresponding to $\Rng{a}$, then the last $a$ in that range (i.e., $L[7]$) is in a marked position since it is the last symbol of a BWT-run, and its position $p_2=4$ in $s$ is precisely the position in $s$ of the last row of the range $\Rng{aa}$. 
For $j=1$, we know that there is an order preserving bijection between the row in $\Rng{aa}=[6,3]$ ending with $b$ and the row in $\Rng{baa}=[1,1]$. Since $x_3=b$ \emph{is} the last symbol in the portion of $L$ corresponding to $\Rng{aa}$ (i.e. $L[8]$), then its position in $s$ is $p_1=p_2-1=4-1=3$, which is also the position of the last, although unique, row in the range $\Rng{baa}$.

\end{example}

\ignore{
\begin{figure}[tb!]
{\small
$$
\begin{array}{cc@{\ }c@{\ }c@{\ }c@{\ }c@{\ }c@{\ }c@{\ }c@{\ }c@{\ }c@{\ }c@{\ }l@{\ }}
F  &  &  &  &  &  &  &  &  &  &  &  &  &  & L \\
\downarrow  &  &  &  &  &  &  &  &  &  &  &  &  &  & \downarrow \\
a & a & a & b & a & a & a & b & a & c & a & b & a & a & b \\
a & a & a & b & a & c & a & b & a & a & b & a & a & a & b \\
\textbf{a} & \textbf{a} & \textbf{b} & \textbf{a} & \textbf{a} & \textbf{a} & \textbf{b} & \textbf{a} & \textbf{a} & \textbf{a} & \textbf{b} & \textbf{a} & \textbf{c} & \textbf{a} & \textbf{b} \\
a & a & b & a & a & a & b & a & c & a & b & a & a & b & a \\
\textbf{a} & \textbf{a} & \textbf{b} & \textbf{a} & \textbf{c} & \textbf{a} & \textbf{b} & \textbf{a} & \textbf{a} & \textbf{b} & \textbf{a} & \textbf{a} & \textbf{a} & \textbf{b} & \textbf{a} \\
\textbf{a} & \textbf{b} & \textbf{a} & {\color{blue} \textbf{c}} & \textbf{a} & \textbf{b} & \textbf{a} & \textbf{a} & \textbf{b} & \textbf{a} & \textbf{a} & \textbf{a} & \textbf{b} & \textbf{a} & \textbf{a} \\
\textbf{a} & \textbf{b} & \textbf{a} & {\color{blue} \textbf{a}} & \textbf{a} & \textbf{b} & \textbf{a} & \textbf{a} & \textbf{a} & \textbf{b} & \textbf{a} & \textbf{c} & \textbf{a} & \textbf{b} & \textbf{a} \\
a & b & a & {\color{blue} a} & a & b & a & c & a & b & a & a & b & a & a \\
a & b & a & {\color{blue} a} & b & a & a & a & b & a & a & a & b & a & c \\
a & c & a & b & a & a & b & a & a & a & b & a & a & a & b \\
b & a & a & a & b & a & a & a & b & a & c & a & b & a & a \\
b & a & a & a & b & a & c & a & b & a & a & b & a & a & a \\
b & a & a & b & a & a & a & b & a & a & a & b & a & c & a \\
b & a & c & a & b & a & a & b & a & a & a & b & a & a & a \\
c & a & b & a & a & b & a & a & a & b & a & a & a & b & a
\end{array}
$$
}
\caption{
\darkgreen{The generalized \BWT matrix for the string $s=abaabaaabaaabac$ computed using 
the orderings $\pi_{aba} = (b,c,a)$, and $\pi_x = (a,b,c)$ for every other substring $x$.}
}\label{fig:BWTx2}
\end{figure}
}   

In addition to the Toehold Lemma, the only other ingredients of the \rindex are 1) the predecessor data structure $P^\pm$ from Lemma~3.5 in~\cite{GNPjacm19}, which is not related to the BWT, and 2) the property that if two consecutive rows of the BWT matrix start and end with the same symbols, then rotating them rightward by one position yields two new rows which are still consecutive and in the same relative order. This property is valid for local orderings, even if it not valid for the general class of context adaptive alphabet orderings, as shown in the following example. 

\begin{example}
Figure ~\ref{fig:BWTx2} shows a generalised BWT matrix based on context adaptive alphabet orderings (on the left) and a local ordering BWT matrix (on the right) for the string $s=baaabaabaac$. Let us consider the rows $3$ and $4$ of the generalised BWT matrix on the left, both starting with $a$ and ending with $b$, highlighted in cyan and light brown, respectively. It can be verified that, if we rotate them rightward yields rows $11$ and $9$, respectively, which are no longer consecutive nor in the same relative order. Instead, let consider the rows $3$ and $4$ of the local ordering BWT matrix, both starting with $a$ and ending with $b$, highlighted in cyan and light brown, respectively. If we rotate them rightward we obtain the rows $10$ and $11$, respectively, which are consecutive and in the same relative order.  
\end{example}

\begin{figure}[tb]
\begin{minipage}[b]{0.47\linewidth}
 {\small
$$
\begin{array}{cc@{\ \ }c@{\ \ }c@{\ \ }c@{\ \ }c@{\ \ }c@{\ \ }c@{\ \ }c@{\ \ }c@{\ \ }c@{\ \ }c@{\ \ }l@{\ \ }}
 &   F & & & &     & &      &  &      &    & L & \\
 &  \downarrow& & & & &  &&&& &\downarrow &\\
 1&  a & a & a & b & a & a & b & a & a & c & b &\\ 
 2&  a & a & b & a& a & b & a & a & c & b & a & \\
 3& \cellcolor{cyan!40} a & a & b & a & a & c & b & a & a & a & \cellcolor{cyan!40}b &\\
 4& \cellcolor{brown!20}a & a & c & b & a & a & a & b & a & a & \cellcolor{brown!20}b &\\
 5&  a & b & a& a & b & a & a & c & b & a & a & \\ 
 6&  a & b & a & a & c & b & a & a & a & b & a &\\ 
 7&  a & c & b & a & a & a & b & a & a & b & a &\\
 8&  c & b & a & a & a & b & a & a & b & a & a &\\ 
 9&  \cellcolor{brown!20}b & \cellcolor{brown!20}a & a & c & b & a & a & a & b & a & a &\\ 
 10& b & a & a & a & b & a & a & b & a & a & c & \leftarrow s\\ 
 11& \cellcolor{cyan!40}b & \cellcolor{cyan!40}a & a & b & a & a & c & b & a & a & a &
\end{array}
$$
}   
\end{minipage}
\hfill
\begin{minipage}[b]{0.49\linewidth}
{\small
$$
   \begin{array}{cc@{\ \ }c@{\ \ }c@{\ \ }c@{\ \ }c@{\ \ }c@{\ \ }c@{\ \ }c@{\ \ }c@{\ \ }c@{\ \ }c@{\ \ }l@{\ \ }}
 & F & & & &       & &    &  &      &    & L & \\
 &\downarrow& & & & & &&&& &\downarrow &\\
1 & a & a & a & b & a & a & b & a & a & c & b &\\ 
2 & a & a & b & a& a & b & a & a & c & b & a  & \\
3 & \cellcolor{cyan!40}a & a & b & a & a & c & b & a & a & a & \cellcolor{cyan!40}b & \\
4 & \cellcolor{brown!20}a & a & c & b & a & a & a & b & a & a & \cellcolor{brown!20}b  &\\
5 & a & b & a& a & b & a & a & c & b & a & a  &\\ 
6 & a & b & a & a & c & b & a & a & a & b & a &\\ 
7  & a & c & b & a & a & a & b & a & a & b & a &\\
8 & c & b & a & a & a & b & a & a & b & a & a &\\
9   & b & a & a & a & b & a & a & b & a & a & c & \leftarrow s\\ 
10 & \cellcolor{cyan!40}b & \cellcolor{cyan!40}a & a & b & a & a & c & b & a & a & a &\\
11 & \cellcolor{brown!20}b &\cellcolor{brown!20} a & a & c & b & a & a & a & b & a & a  
\end{array}
$$
}
 
\end{minipage}

\caption{The generalized \BWT matrix for the string $s=baaabaabaac$  (left) computed using the orderings $\pi_{\epsilon}=(a,c,b)$,  $\pi_{baa} = (c,a,b)$, and $\pi_x = (a,b,c)$ for every other substring $x$. The local ordering BWT matrix for the same string (right) using the orderings $\pi_{\epsilon}=(a,c,b)$, $\pi_{a} = \pi_b= \pi_c= (a,b,c)$. 
In both matrices the horizontal arrow marks the position of the original string $s$, and the last column $L$ is the output of the transformation.}
\label{fig:BWTx2}
\end{figure}

For the local orderings with $k>1$ the above arguments can be generalized with the limitation that the length of the searched pattern must be at least~$k$. The main modification is that we need to store the range $R[y]$ for all strings $y$ of length~$k$ that occur in~$s$. For each such range we also store the suffix array sample for the last row in the range for an overall extra space of $\Oh(\asize^k)$. To search a pattern $x = x_1 \cdots x_p$ we consider the length-$k$ suffix $x'=x_{p-k+1} \cdots x_p$ and we retrieve its range $R[x']$ and the position in $s$ of the last row in the range. With this information we then use Lemma~\ref{lem:rindex} to retrieve in $p-k$ steps the range $R[x_1 \cdots x_p]$ and the position in $s$ of the last row of such range.

\subsection{An alternative view of local orderings}\label{sec:alt}

We conclude this section showing an alternative way to derive transformations based on local orderings. Consider for simplicity the case $k=1$ and assume the transformation $\BWTx$ is defined by the $\sigma+1$ orderings $\pi_\epsilon$ and $\pi_c$ for $c\in\Alpha$.  Consider now the ordering $\Pi$ over $\Alpha^2 = \Alpha \times \Alpha$ defined as follows: Given the pairs $x_1x_2$, $y_1y_2$ in $\Alpha^2$ it is $(x_1x_2 <_\Pi y_1y_2)$ iff $(x_1\neq y_1,\; x_1 <_{\pi_\epsilon} y_1)$, or $(x_1\!=y_1\!=c,\; x_2 <_{\pi_c} y_2)$. We now show that $\BWTx$ is equivalent to the original \BWT over the alphabet $\Alpha^2$ ordered according to~$\Pi$. To each string $s$, we associate a new string $S$ over $\Alpha^2$,  defined by $S[i] = s[i]s[i+1]$ with indices taken modulo $|s|$. Such an approach is described in Example \ref{ex:alternative_local} in which the strings $s=aabaaabac$ and $S= aa\, ab\, ba\, aa\, aa\, ab\, ba\, ac\, ca$ are considered.

\begin{figure}[htb]
{
{\small
$$
\begin{array}{cc@{\ \ }c@{\ \ }c@{\ \ }c@{\ \ }c@{\ \ }c@{\ \ }c@{\ \ }c@{\ \ }c@{\ \ }l@{\ \ }}
 & \multicolumn{9}{c}{M(S)} \\
  &  F  &   &    &    &    &     &   &    & {L}         & \\
  &  \downarrow  &    &    &     &  & & & & \downarrow  & \\
1 & ba & aa & aa & ab & ba & ac & ca & aa & \textcolor{red}{a}b & \\
2 & ba & ac & ca & aa & ab & ba & aa & aa & \textcolor{red}{a}b & \\
3 & ca & aa & ab & ba & aa & aa & ab & ba & \textcolor{red}{a}c & \\
4 & ab & ba & aa & aa & ab & ba & ac & ca & \textcolor{red}{a}a & \\
5 & ab & ba & ac & ca & aa & ab & ba & aa & \textcolor{red}{a}a & \\
6 & aa & ab & ba & aa & aa & ab & ba & ac & \textcolor{red}{c}a & \leftarrow S\\
7 & aa & ab & ba & ac & ca & aa & ab & ba & \textcolor{red}{a}a & \\
8 & aa & aa & ab & ba & ac & ca & aa & ab & \textcolor{red}{b}a & \\
9 & ac & ca & aa & ab & ba & aa & aa & ab & \textcolor{red}{b}a & \\
\end{array}
$$
}
}

\caption{
The \BWT matrix for the string $S=aa\, ab\, ba\, aa\, aa\, ab\, ba\, ac\, ca$ computed using the ordering $\Pi$ based on the local orderings $\pi_\epsilon = (b,c,a)$, $\pi_a = (b,a,c)$, $\pi_b = \pi_c = (a,b,c)$. The concatenation of the pairs in the last column of the matrix gives $bwt(S)=ab\, ab\, ac\, aa\, aa\, ca\, aa\, ba\, ba$. The string obtained by concatenating the first symbol of each pair in $bwt(S)=L$ (coloured in red) is $\BWT_*(s)=aaaaacabb$, where $s=aabaaabac$, as shown in Fig.~\ref{fig:BWTloc}. 
}
\label{fig:BWTlocAltView}
\end{figure}

 There is a natural correspondence between rotations of $s$ and $S$, and because of the definition of $\Pi$, the ordering of $s$'s rotations in $\Mx(s)$ coincides with the ordering of the corresponding rotations of $S$ in $M(S)$. As a consequence, if $\bwt(S)$ (the last column of $M(S)$) has the form $\bwt(S) = x_1y_1\,x_2y_2\cdots x_ny_n$, we have that $\BWTx(s) = x_1 x_2 \cdots x_n$, and the first column of $\Mx(s)$ is $y_1 y_2 \cdots y_n$.
The LF-map applied to $M(S)$ establishes an order preserving bijection between rows ending with $\alpha \in \Alpha^2$ and rows starting with $\alpha$. If $\alpha = x_1x_2$, this translates in $\Mx(s)$ to  an order preserving bijection between rows starting with $x_2$ and ending with $x_1$ and rows starting with $x_1 x_2$: this establishes an alternative proof of Lemma~\ref{lemma:almostbwt}.

The above alternative view of local orderings has probably no practical interest: there is no need to work with the alphabet $\Alpha^2$ to emulate something we can easily do working over $\Alpha$. However, from the theoretical point of view, it is intriguing, and deserving further investigation, that a family of \BWT variants can be obtained by first transforming the string and the alphabet and then applying the standard \BWT followed by the string back-transformation.

\begin{example}\label{ex:alternative_local}
 Let us consider the string $s=aabaaabac$. The local ordering \BWT matrix for $s$ computed using the ordering $\pi_\epsilon = (b,c,a)$, $\pi_a = (b,a,c)$, $\pi_b = \pi_c = (a,b,c)$ is reported in Fig.~\ref{fig:BWTloc}. Instead, Fig.~\ref{fig:BWTlocAltView} shows the BWT matrix $M(S)$ of the string $S= aa\, ab\, ba\, aa\, aa\, ab\, ba\, ac\, ca$ by using the ordering $\Pi$ over the alphabet 
 $\{aa, ab, ac, ba, bb, bc, ca, cb, cc\}$. The last column of $M(S)$ is $bwt(S)=ab\, ab\, ac\, aa\, aa\, ca\, aa\, ba\, ba$. One can verify that there is correspondence between the rows of the local ordering BWT matrix described in Fig. \ref{fig:BWTloc} and the rows of $M(S)$. Moreover, the string obtained by concatenating the first symbol (coloured in red) of each element in $bwt(S)$ is $BWT_*(s)=aaaaacabb$. Finally the string obtained by concatenating the last symbol of each element in $bwt(S)$ gives the first column of the local ordering BWT matrix. 
\end{example}

\section{Run minimization problem}\label{sec:runs}

We consider the following problem: given a string $s$ and a class of \BWT variants, find the variant that minimizes the number of runs in the transformed string. As we mentioned in the introduction, this problem is relevant for the compression of highly repetitive collections. Depending on the class of BWT variants, finding the exact minimum could be a difficult problem, so one may want to resort to heuristics. In this context, any lower bound to the minimum number of runs achievable by a class of transformations would be useful to assess the quality of the solutions found.

In this section we describe an efficient algorithm to determine a lower bound on the number of runs in the transformed string which is valid for all BWT variants discussed in this paper. The lower bound is established by computing the minimal number of runs for the most general class we considered: the class of context adaptive \BWTs described in Section~\ref{sec:genBWT}. In this class we can select an alphabet ordering $\pi_x$ independently for every substring~$x$. It is easy to see that the only orderings that influence the output of the transform are those  associated to strings corresponding to the internal nodes of the suffix tree of~$s$. Nevertheless, the number of possible choices is $\approx {(\sigma!)}^{\Oh(|s|)}$ which is exponential even for constant alphabets.

Given a suffix tree node $v$, we denote by $\bw(v)$ the multiset of symbols associated to the leaves in the subtree rooted at~$v$ (see Fig.~\ref{fig:suffixTree}). We say that a string $z_v$ is a {\em feasible} arrangement of $\bw(v)$ if we can reorder the nodes in the subtree rooted at $v$ so that $z_v$ is obtained by reading left to right the symbols in the reordered subtree. 
For example, in the suffix tree of Figure~\ref{fig:suffixTree} (left), if $v$ is the internal node with upward path $aa$, it is $\bw(v) = \{a,b,c\}$ and $bac$, $bca$, $acb$, $cab$ are feasible arrangements of $\bw(v)$, while $abc$ and $cba$ are {\em not} feasible arrangements.
If $\tau$ is the suffix tree root, using the above notation our problem becomes that of finding the feasible arrangement of $\bw(\tau)$ with the minimal number of runs. The following theorem shows that, for constant alphabets, the optimal arrangement can be found in linear time using dynamic programming. 

\begin{theorem}\label{theo:runs}
Given a string $s$ over an alphabet of size $\sigma = \Oh(1)$, the context adaptive transformation minimizing the number of runs in $\BWTx(s)$ can be found in $\Oh(|s|)$ time.
\end{theorem}

\begin{proof}
Let $\runmin$ denote the minimal number of runs. We show how to compute $\runmin$ with a dynamic programming algorithm; the computation of the alphabet orderings giving $\runmin$ is done using standard techniques. For each suffix tree node $v$ and pairs of symbols $c_i$, $c_j$ let $\Run(v,c_i,c_j)$ denote the minimal number of runs among all feasible arrangements of $\bw(v)$ starting with $c_i$ and ending with $c_j$. Clearly, if $\tau$ is the suffix tree root, then $\runmin = \min_{i,j}\Run(\tau,c_i,c_j)$.

For each leaf $\ell$, it is $\Run(\ell,c_i,c_j)=1$ if $c_i=c_j=\bw(\ell)$ and  $\Run(\ell,c_i,c_j)=\infty$ otherwise. We need to show how to compute, for each internal node $v$, the $\sigma^2$ values $\Run(v,c_i,c_j)$ for $c_i$, $c_j$ in $\Alpha$, given the, up to $\sigma^3$ values, $\Run(w_k,c_\ell,c_m)$, $k=1,\ldots,h$, where  $w_1, \ldots, w_h$ are the children of $v$. To this end, we show that for each ordering $\pi$ of $w_1, \ldots,w_h$ we can compute in constant time the minimal number of runs among all the feasible arrangements of $\bw(v)$ starting with $c_i$ and ending with $c_j$ and with the additional constraint that $v$'s children are ordered according to $\pi$. 

To simplify the notation, assume $w_1,\ldots, w_h$ have been already reordered according to $\pi$. For $k=1,\ldots,h$, let $M_\pi[k,c_\ell,c_m]$ denote the minimal number of runs among all strings $x$ such that $x = y_1 \cdots y_k$, where $y_t$, for $t=1,\ldots,k$, is a feasible arrangement of $\bw(w_t)$, and with the additional constraints that $y_1$ starts with $c_\ell$ and $y_k$ ends with $c_m$ (by construction every such $x$ is a feasible arrangement of~$\bw(v)$)). We have
$$
M_\pi[1,c_\ell,c_m] = \Run(w_1,c_\ell,c_m)
$$
and for $k=2,\ldots,h$
\begin{equation}\label{eq:Mpi}
M_\pi[k,c_\ell,c_m] = \min_{i,j} \left(M_\pi[k-1,c_\ell,c_i]+\Run(w_k,c_j,c_m)-\delta_{ij}\right)
\end{equation}
where $\delta_{ij}=1$ if $i=j$ and $0$ otherwise. Essentially, \eqref{eq:Mpi} states that to find the minimal number of runs for $w_1, \ldots,w_k$ we consider all possible ways to combine an optimal solution for $w_1, \ldots,w_{k-1}$ followed by a feasible arrangement of $\bw({w_k})$. The $\delta_{ij}$ term comes from the fact that the number of runs in the concatenation of two strings is equal to the sum of the runs in each string, minus one if the last symbol of the first string is equal to the first symbol of the second string. 
Once we have the values $M_\pi[h,c_i,c_j]$, the desired values $\Run(v,c_i,c_j)$ are obtained taking the minimum over all possible alphabet ordering~$\pi$.
\end{proof}

\begin{figure}[htb]
\begin{minipage}[b]{0.52\linewidth}
\centering
\includegraphics[scale=0.65]{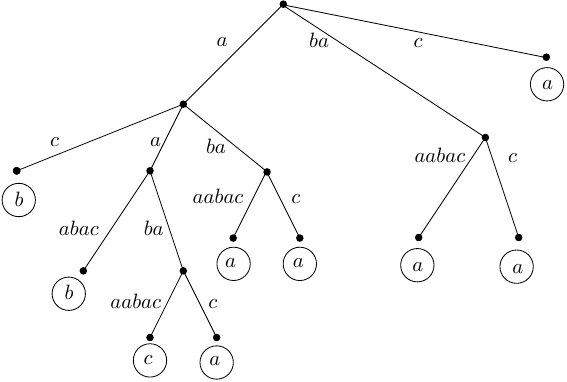}
\end{minipage}
\hfill
\begin{minipage}[b]{0.42\linewidth}
\centering
{\small
$$
\begin{array}{cc@{\ \ }c@{\ \ }c@{\ \ }c@{\ \ }c@{\ \ }c@{\ \ }c@{\ \ }c@{\ \ }c@{\ \ }l@{\ \ }}
 & F & & & &          &  &      &    & L & \\
 &\downarrow& & & &&&& &\downarrow & \\
1 & a & c & a & a & b & a & a & a & b & \\ 
2 & a & a & a & b & a & c & a & a & b & \\ 
3 & a & a & b & a & a & a & b & a & c & \leftarrow s\\ 
4 & a & a & b & a & c & a & a & b & a & \\ 
5 & a & b & a & a& a & b & a & c & a & \\ 
6 & a & b & a & c & a & a & b & a & a & \\
7 & b & a & a & a & b & a & c & a & a & \\ 
8 & b & a & c & a & a & b & a & a & a &  \\ 
9 & c & a & a & b & a & a & a & b & a &  
\end{array}
$$
}
\end{minipage}
\caption{Suffix tree for $s=aabaaabac$, with edges reordered using the run minimization algorithm described in Theorem \ref{theo:runs} (left), The generalized \BWT matrix for the string $s$ computed using the orderings $\pi_a = (c,a,b)$, $\pi_x = (a,b,c)$ for every other substring $x$. The horizontal arrow marks the position of $s$ (right). To each leaf of the tree it is associated the symbol preceding in $s$ the suffix spelled by that leaf. Note that reading left to right the symbols associated to each leaf gives $\BWTx(s)$.}
\label{fig:min_run}
\end{figure}

\begin{example}
The run minimization algorithm described in Theorem \ref{theo:runs} is applied to the standard suffix tree, depicted in Fig. \ref{fig:suffixTree} (left), for the string $s=aabaaabac$, where the symbol $c$ used as a string terminator. In Fig. \ref{fig:min_run} (left) the new tree is shown. It is obtained from the standard suffix tree of $s$ by reordering the children of the internal node with upward path $a$. Denote such a node by $v$.  It is easy to see that $bw(v)=\{a, a, a, b, b, c\}$ and that $bbcaaa$ is a feasible arrangement of $bw(v)$, obtained by moving the third child (with upward path $ac$) to the left. Leaving all other edges unchanged, it is easy to verify that the number of runs remains minimized. The ordering $\pi$ so obtained is defined as $\pi_a=(c,a,b)$ and $\pi_x=(a,b,c)$ for every other substring $x$. The generalized \BWT matrix of the correspondent context adaptive \BWT which minimizes the number of runs is described in Fig. \ref{fig:min_run} (right). 
\end{example}

\section{Conclusions and future directions of research} 

In this paper we introduced a new class of string transformations and showed that they have the same remarkable properties of the \BWT: they can be computed and inverted in linear time, they support linear time pattern search directly in the compressed text, and they can transform a zeroth order compressor into a k-th order compressor (``compression boosting'' property). This implies that such transformations can replace the \BWT in the design of self-indices without any asymptotic loss of performance. Given the crucial role played by the \BWT even outside the area of string algorithms, we believe that expanding the number of efficient \BWT variants can lead to theoretical and practical advancements. A natural consequence will be the design of ``personalized'' transformations, where one will choose the ``best'' alternative to the \BWT according to costs and benefits dictated by application domains. As an example, motivated by the problem of compressing highly repetitive string collections that arises in areas such as bioinformatics, we considered the problem of determining the \BWT variant that minimizes the number of runs in the transformed string. 

Our efficient \BWT variants are a special case of a more general class of transformations that have the same properties of the \BWT but for which we could not devise efficient (linear time) inversion and search algorithms. We believe this larger class of transformation should be further investigated: we have shown that some of them do have more efficient inversion and search algorithms and this suggests that there could be other subclasses of practical interest.  Another possible avenue of further research would be the generalizations of our variants to the recently introduced extension of the BWT in the areas of graphs, languages and automata~\cite{iandc/AlankoDPP21,tcs/GagieMS17,soda/CotumaccioP21}.

\bibliography{BWT}

\begin{thebibliography}{10}

\bibitem{iandc/AlankoDPP21}
Jarno Alanko, Giovanna D'Agostino, Alberto Policriti, and Nicola Prezza.
\newblock Wheeler languages.
\newblock {\em Inf. Comput.}, 281:104820, 2021.

\bibitem{tcs/BannaiGI20}
Hideo Bannai, Travis Gagie, and Tomohiro I.
\newblock Refining the \emph{r}-index.
\newblock {\em Theor. Comput. Sci.}, 812:96--108, 2020.

\bibitem{cpm/BannaiKKP19}
Hideo Bannai, Juha K{\"{a}}rkk{\"{a}}inen, Dominik K{\"{o}}ppl, and Marcin
  Piatkowski.
\newblock Indexing the bijective {BWT}.
\newblock In Nadia Pisanti and Solon~P. Pissis, editors, {\em {CPM}}, volume
  128 of {\em LIPIcs}, pages 17:1--17:14. Schloss Dagstuhl - Leibniz-Zentrum
  f{\"{u}}r Informatik, 2019.

\bibitem{cpm/BannaiKKP21}
Hideo Bannai, Juha K{\"{a}}rkk{\"{a}}inen, Dominik K{\"{o}}ppl, and Marcin
  Piatkowski.
\newblock Constructing the bijective and the extended {Burrows-Wheeler}
  transform in linear time.
\newblock In {\em {CPM}}, volume 191 of {\em LIPIcs}, pages 7:1--7:16,
  Dagstuhl, Germany, 2021. Schloss Dagstuhl -- Leibniz-Zentrum f{\"u}r
  Informatik.

\bibitem{BauerCoxRosoneCPM11}
M.~J. Bauer, A.~J. Cox, and G.~Rosone.
\newblock Lightweight {BWT} construction for very large string collections.
\newblock In {\em CPM}, volume 6661 of {\em LNCS}, pages 219--231, Berlin,
  Heidelberg, 2011. Springer Berlin Heidelberg.

\bibitem{BauerCoxRosoneTCS2013}
M.~J. Bauer, A.~J. Cox, and G.~Rosone.
\newblock Lightweight algorithms for constructing and inverting the {BWT} of
  string collections.
\newblock {\em Theor. Comput. Sci.}, 483(0):134 -- 148, 2013.
\newblock Availability: Code is part of the BEETL library, available as a
  github repository at \url{https://github.com/BEETL/BEETL}.

\bibitem{BNtalg14}
D.~Belazzougui and G.~Navarro.
\newblock Optimal lower and upper bounds for representing sequences.
\newblock {\em ACM T. Algorithms}, 11(4):31:1--31:21, 2015.

\bibitem{BentleyGT19}
Jason~W. Bentley, Daniel Gibney, and Sharma~V. Thankachan.
\newblock On the complexity of {BWT}-runs minimization via alphabet reordering.
\newblock {\em CoRR}, abs/1911.03035, 2019.

\bibitem{esa/BentleyGT20}
Jason~W. Bentley, Daniel Gibney, and Sharma~V. Thankachan.
\newblock On the complexity of {BWT}-runs minimization via alphabet reordering.
\newblock In {\em {ESA}}, volume 173 of {\em LIPIcs}, pages 15:1--15:13,
  Dagstuhl, Germany, 2020. Schloss Dagstuhl--Leibniz-Zentrum f{\"u}r
  Informatik.

\bibitem{BCLRS21_ebwt}
C.~Boucher, D.~Cenzato, Z.~Lipt{\'{a}}k, M.~Rossi, and M.~Sciortino.
\newblock {Computing the Original {eBWT} Faster, Simpler, and with Less
  Memory}.
\newblock In {\em {SPIRE}}, volume 12944 of {\em LNCS}, pages 129--142, Cham,
  2021. Springer International Publishing.

\bibitem{BoucherCL0S21}
C.~Boucher, D.~Cenzato, Z.~Lipt{\'{a}}k, M.~Rossi, and M.~Sciortino.
\newblock r-indexing the {eBWT}.
\newblock In {\em {SPIRE}}, volume 12944 of {\em LNCS}, pages 3--12, Cham,
  2021. Springer International Publishing.

\bibitem{bwt94}
M.~Burrows and D.~J. Wheeler.
\newblock A block sorting data compression algorithm.
\newblock Technical report, DIGITAL System Research Center, 1994.

\bibitem{cazauxRivals2019}
Bastien Cazaux and Eric Rivals.
\newblock {Linking BWT and XBW via Aho-Corasick Automaton: Applications to
  Run-Length Encoding}.
\newblock In {\em CPM}, volume 128 of {\em LIPIcs}, pages 24:1--24:20,
  Dagstuhl, Germany, 2019. Schloss Dagstuhl--Leibniz-Zentrum fuer Informatik.

\bibitem{CGLR_DCC2023}
Davide Cenzato, Veronica Guerrini, Zsuzsanna Lipt{\'{a}}k, and Giovanna Rosone.
\newblock Computing the optimal bwt of very large string collections.
\newblock In {\em Data Compression Conference, {DCC} 2023, Snowbird, UT}.
  {IEEE}, 2023.
\newblock To appear.

\bibitem{CenzatoL22}
Davide Cenzato and Zsuzsanna Lipt{\'{a}}k.
\newblock A theoretical and experimental analysis of {BWT} variants for string
  collections.
\newblock In {\em {CPM}}, volume 223 of {\em LIPIcs}, pages 25:1--25:18.
  Schloss Dagstuhl - Leibniz-Zentrum f{\"{u}}r Informatik, 2022.

\bibitem{dcc/ChapinT98}
B.~Chapin and S.~Tate.
\newblock Higher compression from the {B}urrows-{W}heeler transform by modified
  sorting.
\newblock In {\em DCC}, page 532, Washington, DC, USA, 1998. {IEEE} Computer
  Society.

\bibitem{ChenFoxLyndon1958}
K.~T. Chen, R.~H. Fox, and R.~C. Lyndon.
\newblock Free differential calculus. {IV}. {T}he quotient groups of the lower
  central series.
\newblock {\em Ann. of Math. (2)}, 68:81--95, 1958.

\bibitem{soda/CotumaccioP21}
Nicola Cotumaccio and Nicola Prezza.
\newblock On indexing and compressing finite automata.
\newblock In {\em {SODA}}, pages 2585--2599, USA, 2021. {SIAM}.

\bibitem{CoxBauerJakobiRosone2012}
A.~J. Cox, M.~J. Bauer, T.~Jakobi, and G.~Rosone.
\newblock {Large-scale compression of genomic sequence databases with the
  Burrows-Wheeler transform}.
\newblock {\em Bioinformatics}, 28(11):1415--1419, 2012.
\newblock Availability: Code is part of the BEETL library, available as a
  github repository at \url{https://github.com/BEETL/BEETL}.

\bibitem{CDP2005}
M.~Crochemore, J.~D\'esarm\'enien, and D.~Perrin.
\newblock A note on the {B}urrows-{W}heeler transformation.
\newblock {\em Theor. Comput. Sci.}, 332:567--572, 2005.

\bibitem{spe/CulpepperPP12}
J.~S. Culpepper, M.~Petri, and S.~J. Puglisi.
\newblock Revisiting bounded context block-sorting transformations.
\newblock {\em Software Pract. Exper.}, 42(8):1037--1054, 2012.

\bibitem{DAYKIN2017}
J.~Daykin, R.~Groult, Y.~Guesnet, T.~Lecroq, A.~Lefebvre, M.~L{\'e}onard, and
  {\'E}.~Prieur-Gaston.
\newblock A survey of string orderings and their application to the
  {Burrows}-{Wheeler} transform.
\newblock {\em Theor. Comput. Sci.}, 2017.

\bibitem{tcs/DaykinMS21}
Jacqueline~W. Daykin, Neerja Mhaskar, and W.~F. Smyth.
\newblock Computation of the suffix array, {Burrows-Wheeler} transform and
  {FM}-index in \emph{V}-order.
\newblock {\em Theor. Comput. Sci.}, 880:82--96, 2021.

\bibitem{DaykinIliopoulosSmyth_1994}
J.W. Daykin, C.S. Iliopoulos, and W.F. Smyth.
\newblock Parallel {RAM} algorithms for factorizing words.
\newblock {\em Theor. Comput. Sci.}, 127(1):53 -- 67, 1994.

\bibitem{tcs/Egidi2020}
Lavinia Egidi and Giovanni Manzini.
\newblock {Lightweight merging of compressed indices based on BWT variants}.
\newblock {\em Theor. Comput. Sci.}, 812:214--229, 2020.

\bibitem{cj/fenwick96}
P.~Fenwick.
\newblock The {B}urrows-{W}heeler transform for block sorting text compression:
  Principles and improvements.
\newblock {\em Comput. J.}, 39(9):731--740, 1996.

\bibitem{FGMS2005}
P.~Ferragina, R.~Giancarlo, G.~Manzini, and M.~Sciortino.
\newblock Boosting textual compression in optimal linear time.
\newblock {\em J. ACM}, 52(4):688--713, 2005.

\bibitem{Ferragina:2000}
P.~Ferragina and G.~Manzini.
\newblock Opportunistic data structures with applications.
\newblock In {\em FOCS 2000}, pages 390--398, Washington, DC, USA, 2000. IEEE
  Computer Society.

\bibitem{Ferragina:2005}
P.~Ferragina and G.~Manzini.
\newblock Indexing compressed text.
\newblock {\em J. ACM}, 52:552--581, 2005.

\bibitem{talg/FerraginaV10}
Paolo Ferragina and Rossano Venturini.
\newblock The compressed permuterm index.
\newblock {\em {ACM} Trans. Algorithms}, 7(1):10:1--10:21, 2010.

\bibitem{FRSU_CPM23}
G.~Fici, G.~Romana, M.~Sciortino, and C.~Urbina.
\newblock {On the Impact of Morphisms on BWT-Runs}.
\newblock In {\em {CPM}}, LIPIcs. Schloss Dagstuhl - Leibniz-Zentrum f{\"{u}}r
  Informatik, 2023.
\newblock To appear.

\bibitem{FrosiniMRRS22}
A.~Frosini, I.~Mancini, S.~Rinaldi, G.~Romana, and M.~Sciortino.
\newblock {Logarithmic Equal-Letter Runs for BWT of Purely Morphic Words}.
\newblock In {\em {DLT}}, volume 13257 of {\em LNCS}, pages 139--151. Springer,
  2022.

\bibitem{tcs/GagieMS17}
T.~Gagie, G.~Manzini, and J.~Sir{\'{e}}n.
\newblock Wheeler graphs: {A} framework for {BWT}-based data structures.
\newblock {\em Theor. Comput. Sci.}, 698:67--78, 2017.

\bibitem{GNPjacm19}
T.~Gagie, G.~Navarro, and N.~Prezza.
\newblock Fully-functional suffix trees and optimal text searching in
  {BWT}-runs bounded space.
\newblock {\em Journal of the ACM}, 67(1):article 2, 2020.

\bibitem{GesselRestivoReutenauer2012}
I.~M. Gessel, A.~Restivo, and C.~Reutenauer.
\newblock A bijection between words and multisets of necklaces.
\newblock {\em Eur. J. Combin.}, 33(7):1537 -- 1546, 2012.

\bibitem{GeRe}
I.~M. Gessel and C.~Reutenauer.
\newblock Counting permutations with given cycle structure and descent set.
\newblock {\em J. Comb. Theory A}, 64(2):189--215, 1993.

\bibitem{dlt/GiancarloMRRS18}
R.~Giancarlo, G.~Manzini, A.~Restivo, G.~Rosone, and M.~Sciortino.
\newblock Block sorting-based transformations on words: Beyond the magic {BWT}.
\newblock In {\em {DLT}}, volume 11088 of {\em LNCS}, pages 1--17, Cham, 2018.
  Springer.

\bibitem{cpm/GiancarloMRS19}
R:~Giancarlo, G.~Manzini, G.~Rosone, and M.~Sciortino.
\newblock A new class of searchable and provably highly compressible string
  transformations.
\newblock In {\em {CPM}}, volume 128 of {\em LIPIcs}, pages 12:1--12:12,
  Dagstuhl, Germany, 2019. Schloss Dagstuhl - Leibniz-Zentrum f{\"{u}}r
  Informatik.

\bibitem{GIA07}
R.~Giancarlo, A.~Restivo, and M.~Sciortino.
\newblock From first principles to the {B}urrows and {W}heeler transform and
  beyond, via combinatorial optimization.
\newblock {\em Theor. Comput. Sci.}, 387:236 -- 248, 2007.

\bibitem{abwt_jou}
Raffaele Giancarlo, Giovanni Manzini, Antonio Restivo, Giovanna Rosone, and
  Marinella Sciortino.
\newblock The alternating {BWT}: An algorithmic perspective.
\newblock {\em Theor. Comput. Sci.}, 812:230--243, 2020.

\bibitem{GilScottArxiv2012}
J.~Y. Gil and D.~A. Scott.
\newblock A bijective string sorting transform.
\newblock {\em CoRR}, abs/1201.3077, 2012.

\bibitem{GuerriniLR22}
Veronica Guerrini., Felipe~A. Louza., and Giovanna Rosone.
\newblock Lossy compressor preserving variant calling through extended {BWT}.
\newblock In {\em BIOSTEC/BIOINFORMATICS}, pages 38--48. INSTICC, SciTePress,
  2022.

\bibitem{cpm/HonKLST12}
Wing{-}Kai Hon, Tsung{-}Han Ku, Chen{-}Hua Lu, Rahul Shah, and Sharma~V.
  Thankachan.
\newblock {Efficient Algorithm for Circular Burrows-Wheeler Transform}.
\newblock In {\em {CPM}}, volume 7354 of {\em LNCS}, pages 257--268, Berlin,
  Heidelberg, 2012. Springer Berlin Heidelberg.

\bibitem{isaac/HonLST11}
Wing{-}Kai Hon, Chen{-}Hua Lu, Rahul Shah, and Sharma~V. Thankachan.
\newblock Succinct indexes for circular patterns.
\newblock In {\em {ISAAC}}, volume 7074 of {\em LNCS}, pages 673--682, Berlin,
  Heidelberg, 2011. Springer Berlin Heidelberg.

\bibitem{KaplanVerbin07}
H.~Kaplan and E.~Verbin.
\newblock Most {B}urrows--{W}heeler based compressors are not optimal.
\newblock In {\em CPM}, volume 4580 of {\em LNCS}, pages 107--118, Berlin,
  Heidelberg, 2007. Springer Berlin Heidelberg.

\bibitem{koma00}
R.~Kosaraju and G.~Manzini.
\newblock Compression of low entropy strings with {Lempel--Ziv} algorithms.
\newblock {\em SIAM J. Comput.}, 29(3):893--911, 1999.

\bibitem{Li2014ropebwt}
Heng Li.
\newblock {Fast construction of FM-index for long sequence reads}.
\newblock {\em Bioinformatics}, 30(22):3274--3275, 2014.
\newblock Source code: \url{https://github.com/lh3/ropebwt2}.

\bibitem{Mak15}
V.~M{\"a}kinen, D.~Belazzougui, F.~Cunial, and A.~Tomescu.
\newblock {\em Genome-Scale Algorithm Design}.
\newblock Cambridge University Press, Cambridge, U.K., 2015.
\newblock ISBN 978-1-107-07853-6.

\bibitem{jcb/MakinenNSV10}
V.~M{\"{a}}kinen, G.~Navarro, J.~Sir{\'{e}}n, and Niko V{\"{a}}lim{\"{a}}ki.
\newblock Storage and retrieval of highly repetitive sequence collections.
\newblock {\em J. Comput. Biol.}, 17(3):281--308, 2010.

\bibitem{MantaciRRS07}
S.~Mantaci, A.~Restivo, G.~Rosone, and M.~Sciortino.
\newblock An extension of the {B}urrows-{W}heeler {T}ransform.
\newblock {\em Theor. Comput. Sci.}, 387(3):298--312, 2007.

\bibitem{MantaciRRSV17}
S.~Mantaci, A.~Restivo, G.~Rosone, M.~Sciortino, and L.~Versari.
\newblock Measuring the clustering effect of {BWT} via {RLE}.
\newblock {\em Theor. Comput. Sci.}, 698:79--87, 2017.

\bibitem{MantaciRRS17}
Sabrina Mantaci, Antonio Restivo, Giovanna Rosone, and Marinella Sciortino.
\newblock {Burrows-Wheeler Transform and Run-Length Enconding}.
\newblock In {\em {WORDS}}, volume 10432 of {\em LNCS}, pages 228--239.
  Springer, 2017.

\bibitem{Manzini2001}
G.~Manzini.
\newblock An analysis of the {B}urrows-{W}heeler transform.
\newblock {\em J. ACM}, 48(3):407--430, 2001.

\bibitem{Nav16}
G.~Navarro.
\newblock {\em Compact Data Structures -- A practical approach}.
\newblock Cambridge University Press, Cambridge, U.K., 2016.

\bibitem{NM-survey07}
G.~Navarro and V.~M{\"a}kinen.
\newblock Compressed full-text indexes.
\newblock {\em ACM Comput. Surv.}, 39(1), 2007.

\bibitem{csur/Navarro21a}
Gonzalo Navarro.
\newblock Indexing highly repetitive string collections, part {I:}
  repetitiveness measures.
\newblock {\em {ACM} Comput. Surv.}, 54(2):29:1--29:31, 2021.

\bibitem{csur/Navarro21}
Gonzalo Navarro.
\newblock Indexing highly repetitive string collections, part {II:} compressed
  indexes.
\newblock {\em {ACM} Comput. Surv.}, 54(2):26:1--26:32, 2021.

\bibitem{ccp/PetriNCP11}
M.~Petri, G.~Navarro, J.~S. Culpepper, and S.~J. Puglisi.
\newblock Backwards search in context bound text transformations.
\newblock In {\em {CCP}}, pages 82--91, Washington, DC, USA, 2011. {IEEE}
  Computer Society.

\bibitem{RestivoRosoneTCS2011}
A.~Restivo and G.~Rosone.
\newblock Balancing and clustering of words in the {B}urrows-{W}heeler
  transform.
\newblock {\em Theor. Comput. Sci.}, 412(27):3019 -- 3032, 2011.

\bibitem{Schindler1997}
M.~Schindler.
\newblock A fast block-sorting algorithm for lossless data compression.
\newblock In {\em DCC}, page 469, Washington, DC, USA, 1997. {IEEE} Computer
  Society.

\end{thebibliography}

\end{document}